\documentclass{aa}  
\usepackage{txfonts}
\usepackage{lscape}
\usepackage{longtable}
\usepackage{psfig}
\usepackage{natbib}
\bibpunct{(}{)}{;}{a}{}{,}

\begin{document}
   \title{Lithium abundances of halo dwarfs based on excitation temperatures. II. NLTE}

   \author{A. Hosford\inst{1}
          \and
          A.E. Garc\'{i}a P\'{e}rez\inst{1}
          \and
 					R. Collet\inst{2}
 					\and
          S.G. Ryan\inst{1}
          \and
          J.E. Norris\inst{3}
          \and
          K.A. Olive\inst{4}
          }

   \offprints{A. Hosford}

   \institute{Centre for Astrophysics Research, University of Hertfordshire,
              College Lane, Hatfield, AL10 9AB, UK\\
              \email{a.hosford@herts.ac.uk, a.e.garcia-perez@herts.ac.uk, s.g.ryan@herts.ac.uk}
				 \and
         		 Max-Planck-Institut f$\rm \ddot{u}$r Astrophysik, Postfach 1317, D-85741 Garching bei M$\rm \ddot{u}$nchen, Germany\\
         		 \email{remo@mpa-garching.mpg.de}
         \and
             Research School of Astronomy and Astrophysics, The Australian National University, Mount Stromlo Observatory, 								Cotter Road, Weston, ACT 2611, Australia \\
             \email{jen@mso.anu.edu.au}
         \and
         		William I. Fine Theoretical Physics Institute, School of Physics and Astronomy, University of Minnesota, 			   						Minneapolis, MN 55455, USA\\
         		\email{OLIVE@umn.edu} 
             }

   \date{Received; accepted}

% \abstract{}{}{}{}{} 
% 5 {} token are mandatory
 
  \abstract
   {The plateau in the abundance of \element[][7]{Li} in metal-poor stars was initially interpreted as an observational indicator of the primordial lithium abundance. However, this observational value is in disagreement with that deduced from calculations of  Big Bang nucleosynthesis (BBN), when using the Wilkinson microwave anisotropy probe (WMAP) baryon density measurements. One of the most important factors in determining the stellar lithium abundance is the effective temperature. In a previous study by the authors, new effective temperatures ($T_{\rm eff}$) for sixteen metal-poor halo dwarfs were derived using a local thermodynamic equilibrium (LTE) description of the formation of Fe lines. This new $T_{\rm eff}$ scale reinforced the discrepancy.}
   {For six of the stars from our previous study we calculate revised temperatures using a non-local thermodynamic equilibrium (NLTE) approach. These are then used to derive a new mean primordial lithium abundance in an attempt to solve the lithium discrepancy.}
   {Using the code ${\rm\sc MULTI}$ we calculate NLTE corrections to the LTE abundances for the \ion{Fe}{i} lines measured in the six stars, and determine new $T_{\rm eff}$'s. We keep other physical parameters, i.e. log $g$, [Fe/H] and $\xi$, constant at the values calculated in Paper I. With the revised $T_{\rm eff}$ scale we derive new Li abundances. We compare the NLTE values of $T_{\rm eff}$ with the photometric temperatures of Ryan et al. (1999, ApJ, 523, 654), the infrared flux method (IRFM) temperatures of Mel\'{e}ndez {\&} Ram\'{i}rez (2004, ApJ, 615, 33), and the Balmer line wing temperatures of Asplund et al. (2006, ApJ, 644, 229).}
   {We find that our temperatures are hotter than both the Ryan et al. and Asplund et al. temperatures by typically $\sim$ 110 K - 160 K, but are still cooler than the temperatures of Mel\'{e}ndez {\&} Ram\'{i}rez by typically $\sim$ 190 K. The temperatures imply a primordial Li abundance of 2.19 dex or 2.21 dex, depending on the magnitude of collisions with hydrogen in the calculations, still well below the value of 2.72 dex inferred from WMAP + BBN. We discuss the effects of collisions on trends of \element[][7]{Li} abundances with [Fe/H] and $T_{\rm eff}$, as well as the NLTE effects on the determination of log $g$ through ionization equilibrium, which imply a collisional scaling factor $\rm S_{H} >$ 1 for collisions between \ion{Fe}{} and H atoms.}
   {}


   \maketitle
%
%________________________________________________________________

\section{Introduction}

Since its discovery by \citet{SpiteSpite1982}, many studies of the plateau in lithium in metal-poor dwarfs have been undertaken, e.g. \citet{Spiteetal1996}, \citet{Ryanetal2000}, \citet{MelendezRamirez2004}, \citet{Bonifacioetal2007} and \citet{Aokietal2009}, confirming its existence. Most studies find a comparable Li abundance ($A$(Li)\footnote{$A$(Li)$\equiv log_{10} \left(\frac{N(\rm Li)}{N(\rm H)}\right)+12.00$} $\approx$ 2.0 - 2.1 dex)  yet discrepancies still exist, in particular the high value found by \citet{MelendezRamirez2004} ($A$(Li) = 2.37 dex). However, the biggest discrepancy comes from a comparison of the primordial abundances inferred from observations and that derived from Big Bang Nucleosynthesis (BBN) with the WMAP constraint on the baryon density fraction, $\Omega_{\rm B}h^{2}$, which leads to $A$(Li) = 2.72 dex \citep{Cyburt2008}. This is what has become known as the ``lithium problem''.

Several possibilities have been proposed to explain this discrepancy. Broadly these are: systematic errors in the derived stellar Li abundances; errors in the BBN calculations due to uncertainties in some of the relevant nuclear reaction rates; the destruction of some of the BBN-produced Li prior to the formation of the stars we have observed; the introduction of new physics that may affect BBN \citep{JedamzikPospelov2009} ; or the removal of Li from the photospheres of the stars through their lifetimes \citep[see introduction to][Paper I, for more details]{Hosfordetal2009}. The possible explanation under study in this work is that of systematic errors in the effective temperature ($\textit{T}_{\rm eff}$) scale for metal-poor stars. The effective temperature is the most important atmospheric parameter affecting the determination of Li abundances. This is due to the high sensitivity of $A$(Li) to $\textit{T}_{\rm eff}$, with $\partial{A}/\partial{T_{\rm eff}} {\sim}$ 0.065 dex per 100 K. One reason for the spread in the observed $A$(Li) is the differences in the $\textit{T}_{\rm eff}$ scales used by different authors. For instance,  \citet{Spiteetal1996} and \citet{Asplundetal2006} derive a $\textit{T}_{\rm eff}$ of 5540 K and 5753 K for the star HD140283, respectively. The scale of \citet{MelendezRamirez2004} is on average $\sim$ 200 K hotter than other works. This goes some way to explaining their higher $A$(Li); other factors, such as the model atmospheres with convective overshooting used in their work, may also contribute to the discrepancy. It is important to confirm, or rule out, whether systematic errors in $\textit{T}_{\rm eff}$ are the cause of the Li problem, and in doing so address the need for other possible explanations.

In previous work \citep[- Paper I]{Hosfordetal2009}, we utilised the exponential sensitivity in the Boltzmann distribution to $\chi/T$, where $\chi$ is the excitation energy of the lower level of a transition. Using this, we determined $\textit{T}_{\rm eff}$'s for eighteen metal-poor stars close to the main-sequence turnoff. This was done by nulling the dependence of $A$(Fe) on $\chi$ for approx 80 -- 150 \ion{Fe}{i} lines. Two $\textit{T}_{\rm eff}$ scales were generated due to uncertainty in the evolutionary state of some of the stars under study. It was found that our temperatures were in good agreement with those derived by a Balmer line wing method by \citet{Asplundetal2006} and those derived by photometric techniques by \citet{Ryanetal1999}. However, our $\textit{T}_{\rm eff}$ scale was on average $\sim$ 250 K cooler than temperatures from the infrared flux method (IRFM) as implemented by \citet{MelendezRamirez2004}. This is not the case for all work done using the IRFM, the IRFM effective temperatures of \citet{Alonsoetal1996} are similar to ours, for stars we have in common. 

The derived mean abundances in Paper I were $A$(Li) = 2.16 dex assuming main-sequence (MS) membership and $A$(Li) = 2.10 dex assuming sub-giant branch (SGB) membership. For the five stars that have a known evolutionary state, we calculated a mean $A$(Li) = 2.18 dex. It is clear that these values are not high enough to solve the lithium problem. However, the analysis of \citet{Hosfordetal2009} assumed that the spectrum was formed in local thermodynamic equilibrium (LTE). This is a standard way of calculating spectra, but oversimplifies the radiative transfer problem, and it was acknowledged in \citet{Hosfordetal2009} that LTE simplification affect those results. Consequently, although it was shown that, within the LTE framework, systematic errors in the $\textit{T}_{\rm eff}$ scale are not the cause of the disparity between spectroscopic and BBN+WMAP values for the primordial Li, we also need to assess the impact of non-local thermodynamic equilibrium (NLTE) on the determination of stars effective temperatures. That is the aim of the current work.

This work is not intended to be a full dissection of the methods of NLTE, but rather an application of those more complex (and possibly more accurate) methods to derive a new $\textit{T}_{\rm eff}$ scale and to assess their impact on the lithium problem. However, to do this we need to delve, with some depth, into the processes of NLTE line formation, which we do in Sect. 2. This will give some understanding of the complexities and uncertainties that are involved and give the opportunity to make some generalisations on the important aspects that need to be addressed. In Sect. 3 -- 5 we detail our calculations and results, and discuss these further in Sect. 6.
%__________________________________________________________________

\section{NLTE Framework}
\subsection{The necessity for NLTE}
\label{sec:TheNecessityForNLTE}

With the availability of high quality spectra, the problem of calculating accurate chemical abundances often comes down to a better understanding of the line formation process. This is of particular importance to this work as the calculation of accurate level populations of the \ion{Fe}{i} atom and source functions at the wavelengths of the \ion{Fe}{} transitions is crucial to determining $\textit{T}_{\rm eff}$ from lines of different $\chi$. In LTE calculations, the level populations follow the Boltzmann and Saha distributions. These assume that the levels are populated, or depopulated, by collisional and/or radiative processes, that are characterised by the local kinetic temperature. In the deep layers of the atmosphere, at $\tau_{\rm 5000} > 1$, where $\tau_{\rm 5000}$ is the optical depth at 5000\AA, LTE is a reasonable assumption. However, it tends to break down at optical depths $\tau_{\rm 5000} < 1$, i.e. through most of the line forming region of the photosphere. Therefore neglecting deviations of the level populations from LTE could lead to errors in the $\textit{T}_{\rm eff}$ derived by excitation dependence. Furthermore, in NLTE calculations, it is not only the level populations that differ from the LTE case. The radiative transitions of the atom must be explicitly considered. The fact that the radiation field is no longer described by a Planck function, and certainly not a Planck function calculated for the local temperature, results in further changes of the spectrum relative to the LTE case. This last effect is very important in metal-poor stars, where the reduced opacity/increased transparency of the atmosphere exposes shallow, cooler layers to the UV-rich spectrum coming from the deeper, hotter layers \citep{Asplundetal1999}.

For Fe in particular, different studies have come to different conclusions as to the magnitude of the NLTE corrections. \citet{TandI1999} found that there can be corrections of up to 0.35 dex on \ion{Fe}{i} abundances for main-sequence stars at [Fe/H] $\approx -3$, and suggest that all work done on metal-poor stars should be carried out using NLTE methods. \citet{Grattonetal1999}, however, find negligible corrections to \ion{Fe}{i} abundances and see this as validation that LTE assumptions still hold when studying this type of star. In contrast, work by \citet{Shchulinaetal2005} find higher correction values of $\sim$0.9 dex and $\sim$0.6 dex, depending on whether 3D or 1D atmospheres are used.  The difference in their conclusions is driven principally by the different relative importance of collisional and radiative transitions in their calculations. \citet{Grattonetal1999} have relatively stronger collisional transitions, and as a result find smaller deviations from LTE. \citet{Shchulinaetal2005} include no collisions with neutral hydrogen. We return to this important point below, but for now it illustrates that much work still needs to be done in this field before we can be certain of the impact of NLTE. 

\subsection{The coupling of the radiation field and level populations}
\label{sec:TheCoupingOfTheRadiationFieldAndLevelPopulations}

Many factors have to be taken into account when computing radiative transfer in NLTE. This leads to a complicated situation where, for example, we have to solve population equations and radiative transfer equations simultaneously. This is due to the level populations and the radiation field being coupled, a fact ignored in LTE calculations. There are large uncertainties in NLTE calculations because of the lack of complete information on the rates of collisional and radiative transitions between energy levels for a given element in all its important ionization states. This is especially true for larger atoms which have a greater number of energy levels, as is the case for Fe.

To solve NLTE problems, a system of rate equations is needed that describes fully the populations of each level within the atom under study. Statistical equilibrium is invoked, i.e. the radiation fields and the level populations are constant with time. The formulation of the problem is well described in \citet{Mihalas1978}, from which the following equations are taken. The population of level $i$ is the sum of all the processes that populate the level minus the processes that depopulate it, such that:
\begin{equation}
\frac{{{\rm{d}}n_i }}{{{\rm{d}}t}} = \sum\limits_{j \ne i}^N {n_j P_{ji}  - n_i \sum\limits_{j \ne i}^N {P_{ij}  = 0} } 
\end{equation}
where $\textit{n}_{i}$ and $\textit{n}_{j}$ are the populations of the levels $i$ and $j$ respectively, $N$ is the total number of levels, including continua, and $\textit{P}_{ji}$ and $\textit{P}_{ij}$ are the rates of transitions into and out of the level $i$. The rates are given by:
\begin{equation}
P_{ij}  = A_{ij}  + B_{ij} \bar {J_{\nu _0 } }  + C_{ij} 
\end{equation}
where $\textit{A}_{ij}$, $\textit{B}_{ij}$, and $\textit{C}_{ij}$ are the Einstein coefficients for spontaneous, radiative and collisional excitation respectively, $\nu_{0}$ is the frequency of the transition between levels $i$ and $j$, and $\bar{J}$$_{\nu_{0}}$ is the mean intensity averaged over the line profile. It is usually the case that the radiative rates dominate over the collisional ones at optical depths $\tau_{\rm 5000} <$ 1 implying that LTE assumptions no longer hold in general. The rate equations depend on the mean intensities over the relevant frequencies, $\bar{J}$$_{\nu_{0}}$, meaning that the level populations depend on the radiation field. Conversely the radiation field depends on the level populations through the radiative transfer equation. This is seen by examining the simple problem of the two-level atom where the radiative transfer equation is given by:
\begin{equation}
\mu \frac{{{\rm{d}}I_\nu  }}{{{\rm{d}}\tau_\nu }} = I_\nu   - S^l_\nu 
\end{equation}
where $\textit{I}_{\nu}$ is the intensity at frequency $\nu$, $\mu$ is the cosine of the viewing angle, $\tau_\nu$ is the optical depth and $\textit{S}^{l}_\nu$ is the line source function, such that:
\begin{equation}
S^l_\nu  = \left( {\frac{{2h\nu ^3 }}{{c^2 }}} \right)\left[ {\left( {\frac{{n_i g_j }}{{n_j g_i }}} \right) - 1} \right]^{ - 1} 
\end{equation}
Here $h$ is Planck's constant, $c$ is the speed of light, $\textit{g}_{i}$  and $\textit{g}_{j}$ are the statistical weights of the levels $i$ and $j$ respectively, and $n_i/n_j$ gives the ratio of the populations of the levels $i$ and $j$ calculated using the rate equation, Eq. (1). This form of the source function strictly speaking holds under the assumption of complete frequency redistribution but still illustrates the problem of having to solve the two sets of equations simultaneously.

\subsection{Transition rates}
\label{sec:transitionrates}
For the calculation of the level populations, through Eq. 1, radiative and collisional rates are required.

 For the radiative rates, the bound-bound transition probabilities and photoionization cross sections are needed for all levels of the atom in all significant ionization states. Two of the larger projects providing values for these are the Opacity Project \citep{Seaton1987} and the IRON project \citep{Bautista1997}. For Fe, the Opacity Project finds typically a $>$ 10 $\%$ uncertainty for their photoionization data \citep{Seatonetal1994}. The Bautista photoionization values, which are larger than those previously used, lead to increased photoionization rates \citep{Asplund2005} and hence to lower abundances as overionization becomes more efficient.
 
For the collisional data, large uncertainties still exist. The two main types of collisions that affect the line profile are those with electrons and neutral hydrogen. Coupling of all levels in the Fe model atom occurs due to these types of collisions, especially in the atmospheres of cool stars where electrons and neutral H are believed to be the dominant perturbers. A simple calculation, like that in \citet{Asplund2005}, shows that \ion{H}{I} collisions dominate over electron collisions in thermalizing processes in metal-poor stars and are therefore important in calculations of line profiles. For collisions with neutral hydrogen, the approximate formulation of \citet{Drawin1968, Drawin1969} is used as implemented by \citet{SteenbockHolweger1984}. However, through laboratory testing and quantum calculations of collisions with atoms such as Li and Na, it has been shown that Drawin's formula does not produce the correct order of magnitude result for \ion{H}{I} collisional cross-sections. In some cases, where comparisons with experimental data or theoretical results can be made, the Drawin recipe overestimates the cross-sections by \emph{one} to \emph{six} orders of magnitude \citep[e.g.][]{Flecketal1991, Barklemetal2003}. Corrections to the Drawin cross-sections are suggested by \citet{Lambert1993} to compensate for these differences. 

Due to the uncertainties in the magnitude of the H collisions, the Drawin cross-sections are scaled with a factor $\rm S_{H}$. There are different schools of thought on how to deal with this parameter. \citet{Colletetal2005} treat it as a free parameter in their work, adopting values of $\rm S_{H}$ = 0.001 and 1 and test the effect this has on their results. Higher values of $\rm S_{H}$ correspond to more collisions and hence more LTE-like conditions. Their main aim, however, was to test not the efficiency of H collisions but the effects of line-blocking on the NLTE problem. \citet{Kornetal2003} make it one of their aims to constrain $\rm S_{H}$. To do this, they ensure ionization equilibrium between \ion{Fe}{i} and \ion{Fe}{ii} using the log g derived from {\sc hipparcos} parallax and $\textit{T}_{\rm eff}$ from H lines. In doing this, they find that a value of $\rm S_{H}$ = 3 holds for a group of local metal-poor stars. This apparently contradicts the statement above that Drawin's formula overestimates the cross-sections. \citet{Grattonetal1999} use $\rm S_{H}$ = 30. This value was constrained by increasing  $\rm S_{H}$ until spectral features of several elements, i.e. Fe, O, Na and Mg, of RR Lyrae stars all gave the same abundance. With such elevated collisional rates, \citet{Grattonetal1999} not surprisingly find results very close to LTE, i.e. they find very small NLTE corrections.

Collisions with neutral hydrogen and electrons are important not only in coupling bound states to each other, but also in coupling the whole system to the continuum i.e. to the \ion{Fe}{ii} ground state (and potentially excited states). This is especially true when considering the high excitation levels. These levels are more readily collisionally ionised than lower levels, and are also coupled to each other by low energy (infrared) transitions, therefore thermalization of the levels occurs which drives the populations more towards LTE values. It is therefore important to have a model atom that includes as many of the higher terms of the atom as possible \citep{Korn2008}, although it is not necessary to include all individual levels. We return to this point in Sect. 4. We describe the model atom and calculations next before moving on to the results.

\subsection{The model Fe atom}
\label{sec:TheModelFeAtom}

The Fe model adopted for this work is that of \citet{Colletetal2005}, which is an updated version of the model atom of \citet{TandI1999}. The atom includes 334 levels of \ion{Fe}{i} with the highest level at 6.91 eV. For comparison, the first ionization energy is 7.78 eV and the NIST database lists 493 \ion{Fe}{i} levels. Many of the highest levels are not included in our model; due in part to computational limitations i.e. the more complicated the model, the greater the computer power and time needed to complete the computations, and because of lack of important information, e.g. photoionization cross sections. We report below on the effects the missing upper levels have on the corrections and try to quantify their importance in the NLTE calculations. The model also includes 189 levels of \ion{Fe}{ii} with the highest level at 16.5 eV, and the ground level of \ion{Fe}{iii}. For comparison, the second ionization energy is 16.5 eV, and the NIST database lists 578 \ion{Fe}{ii} levels. This model configuration leads to the possibility of 3466 bound-bound radiative transitions in the \ion{Fe}{i} system, 3440 in the \ion{Fe}{ii} system, and 523 bound-free transitions. We run the calculations with the whole model, but  present results only for the lines that are measured in our program stars.

Oscillator strengths for the \ion{Fe}{i} lines are taken from \citet{Naveetal1994} and \citet{KuruczBell1995}, whilst values from \citet{Fuhretal1988}, \citet{HirataHoraguchi1995}, and \citet{Thevenin1989, Thevenin1990} were used for the \ion{Fe}{ii} lines. The photoionization cross-sections are taken from the IRON Project (Bautista 1997). \citet{Colletetal2005} smoothed these cross-sections so as to minimize the number of wavelength points to speed up the computational processes.

Collisional excitations by electrons are incorporated through the van Regemorter formula \citep{VRegemorter1962} and cross-sections for collisional ionization by electrons are calculated by the methods of \citet{Cox2000}. In the case of H collisions, the approximate description of \citet{Drawin1968, Drawin1969}, as implemented by \citet{SteenbockHolweger1984} with the correction of \citet{Lambert1993} and multiplied by $\rm S_{H}$, has been used. As we do not intend to constrain $\rm S_{H}$, we treat it as a free parameter and adopt values of 0 (no neutral H collisions), 0.001 and 1 (Drawin's prescription). This allows us to assess the importance of H collisions on the NLTE corrections. For all calculations, the oscillator strength value, $f_{ij}$, has been set to a minimum of $10^{-3}$ when there is no reliable data or the $f$ value for a given line is below this minimum. This minimum is set as the scaling between the cross-sections and the $f$ value breaks down for weak and forbidden lines \citep{Lambert1993}.

\subsection{The model atmospheres}
\label{sec:TheModelAtmosphere}

In this work, we have adopted plane-parrallel {\sc MARCS} models. These models are used, rather than the Kurucz 1996 models as was done in \citet{Hosfordetal2009}, as {\sc MULTI} needs a specific format for its input, this is provided by the {\sc MARCS}, details of which can be found in \citet{Asplundetal1997}. 3D models lead to an even steeper temperature gradient, and hence cooler temperatures in the line forming region \citep{Asplund2005}, but the use of these more sophisticated models is beyond the scope of this work.

\subsection{Radiative transfer code}
\label{sec:RadiativeTransferCode}

The NLTE code used to produce Fe line profiles and equivalent widths ($W_{\rm \lambda}$) is a modified version of {\sc MULTI} \citep{Carlsson1986}. This is a multi-level radiative transfer program for solving the statistical equilibrium and radiative transfer equations. The code we adopted is a version modified by R. Collet to include the effects of line-blocking \citep{Colletetal2005}. To do this, they sampled metal line opacities for 9000 wavelength points between 1000 \AA\ and 20000 \AA\ and added them to the standard background continuous opacities. They found that, for metal-poor stars, the difference between NLTE Fe abundances derived from \ion{Fe}{i} lines excluding and including line-blocking by metals in the NLTE calculations is of the order of 0.02 dex or less.

\section{NLTE calculations}
\label{sec:NLTECalculations}

For this work, we have chosen six of our original program stars \citep{Hosfordetal2009} that approximately represent the limits of our physical parameters, i.e. one of the more metal-rich, one of the less metal-rich, one of the hotter, one of the cooler etc. Table \ref{Table1} indicates the stellar parameters for which model atmospheres were created. For the HD stars {\sc hipparcos} gravities were used. For the other three stars, lower and upper limits on log $g$ are given by theoretical isochrones  \citep[see][]{Hosfordetal2009}. In the case of LP815-43, there is uncertainty as to whether it is just above or just below the main-sequence turnoff. The final temperatures are interpolated between these values using a final log $g$ that represents the star at 12.5 Gyr (Table \ref{Table2}). This study is primarily concerned with the formation of \ion{Fe}{i} lines.

\begin{table}[htbp]
\caption{Physical parameters for the atmospheric models used in this work}
	\label{Table1}
	\centering
		\begin{tabular} {l c c c c}
			\hline\hline
			Star	&	$\textit{T}_{\rm eff}$ &	[Fe/H]&	log $g$ &	$\xi$	\\
			    & (K)  &  (dex)   &   (dex)   &   (Km$\rm s^{-1}$) \\
			\hline
			HD140283    & 5769  & $-2.54$ & 3.73  & 1.5 \\
			HD84937     & 6168  & $-2.34$ & 3.98  & 1.3 \\
			HD74000     & 6070  & $-2.20$ & 4.03  & 1.2 \\
			BD+26$^{\circ}$ 2621	&	6225	&	$-2.68$	&	4.47	&	1.2	\\
			BD+26$^{\circ}$ 2621	&	6241	&	$-2.67$	&	4.51	&	1.2	\\
			CD$-$33$^{\circ}$ 1173	&	6380	&	$-2.94$	&	4.41	&	1.5	\\
			CD$-$33$^{\circ}$ 1173	&	6391	&	$-2.94$	&	4.47	&	1.5	\\
			LP815$-$43 (SGB)	&	6383	&	$-2.71$	&	3.80	&	1.4	\\
			LP815$-$43 (SGB)	&	6409	&	$-2.68$	&	3.91	&	1.4	\\
			LP815$-$43 (MS)	&	6515	&	$-2.62$	&	4.35	&	1.4	\\
			LP815$-$43 (MS)	&	6534	&	$-2.61$	&	4.42	&	1.4	\\
			\hline
		\end{tabular}
\end{table}

\begin{figure*}[h]
\centering
\vbox{

\hbox{
	\psfig{file=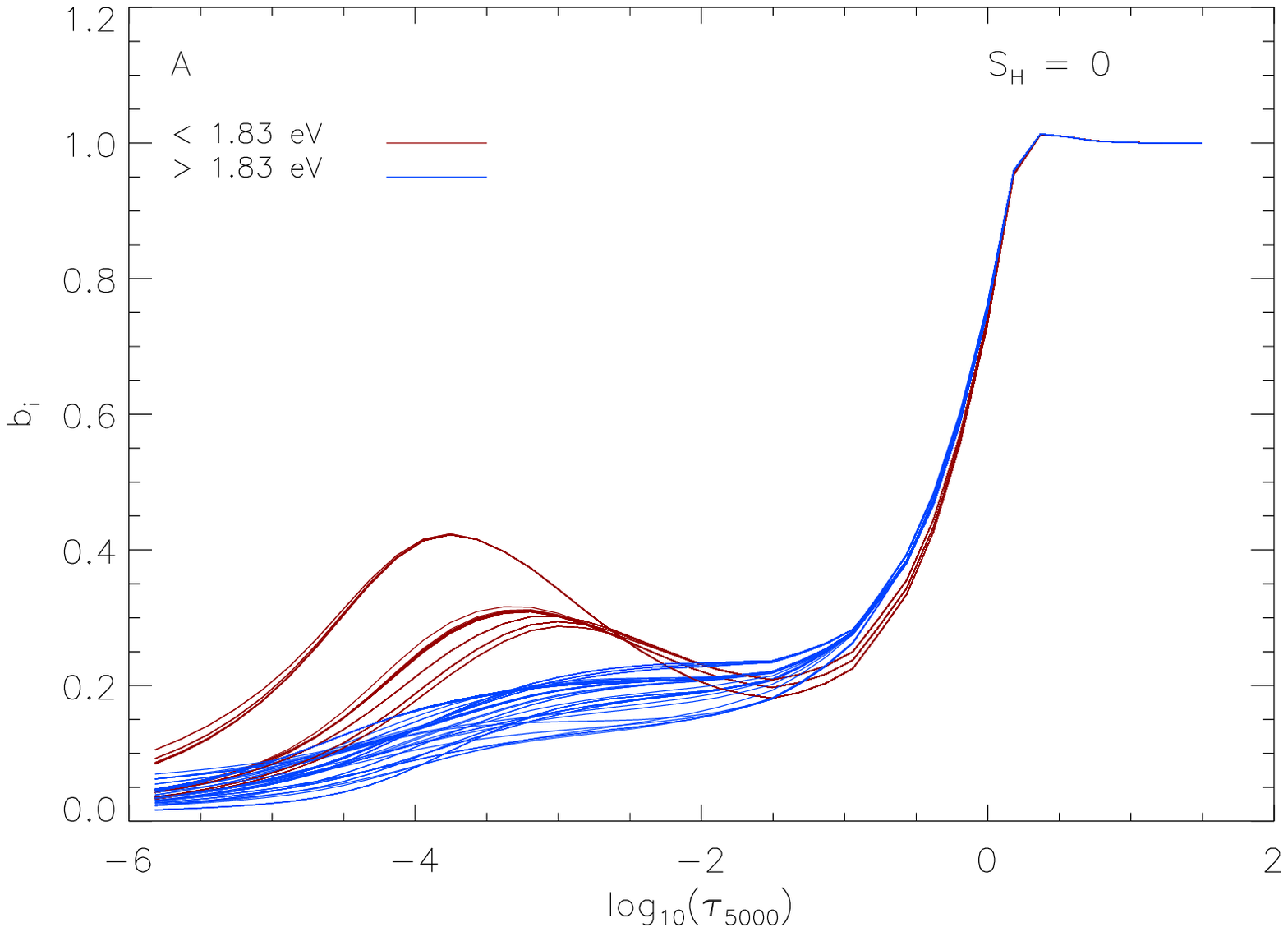,width=9cm,height=5.5cm}
	\psfig{file=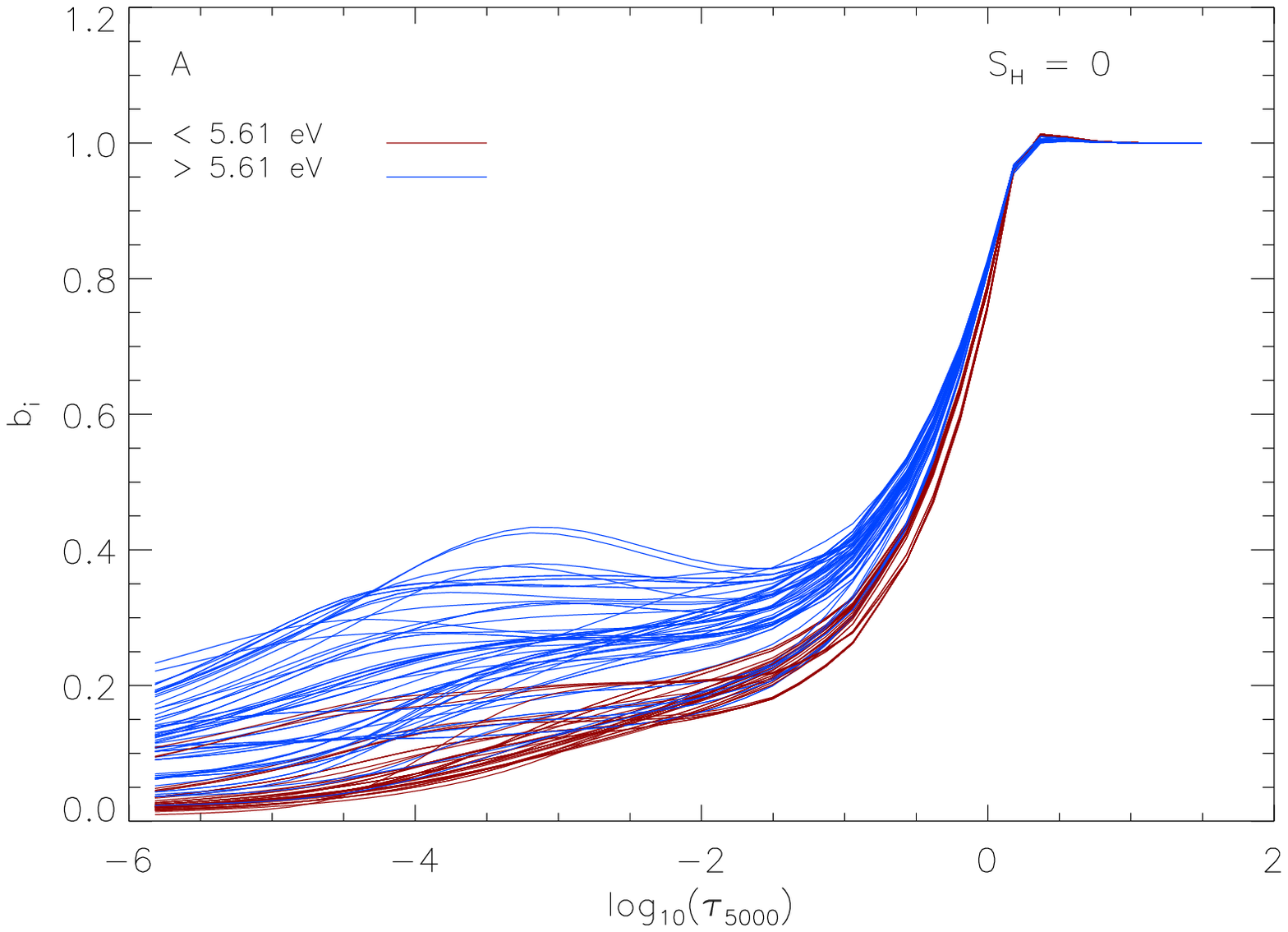,width=9cm,height=5.5cm}}
	}
\vbox{
	\hbox{
	\psfig{file=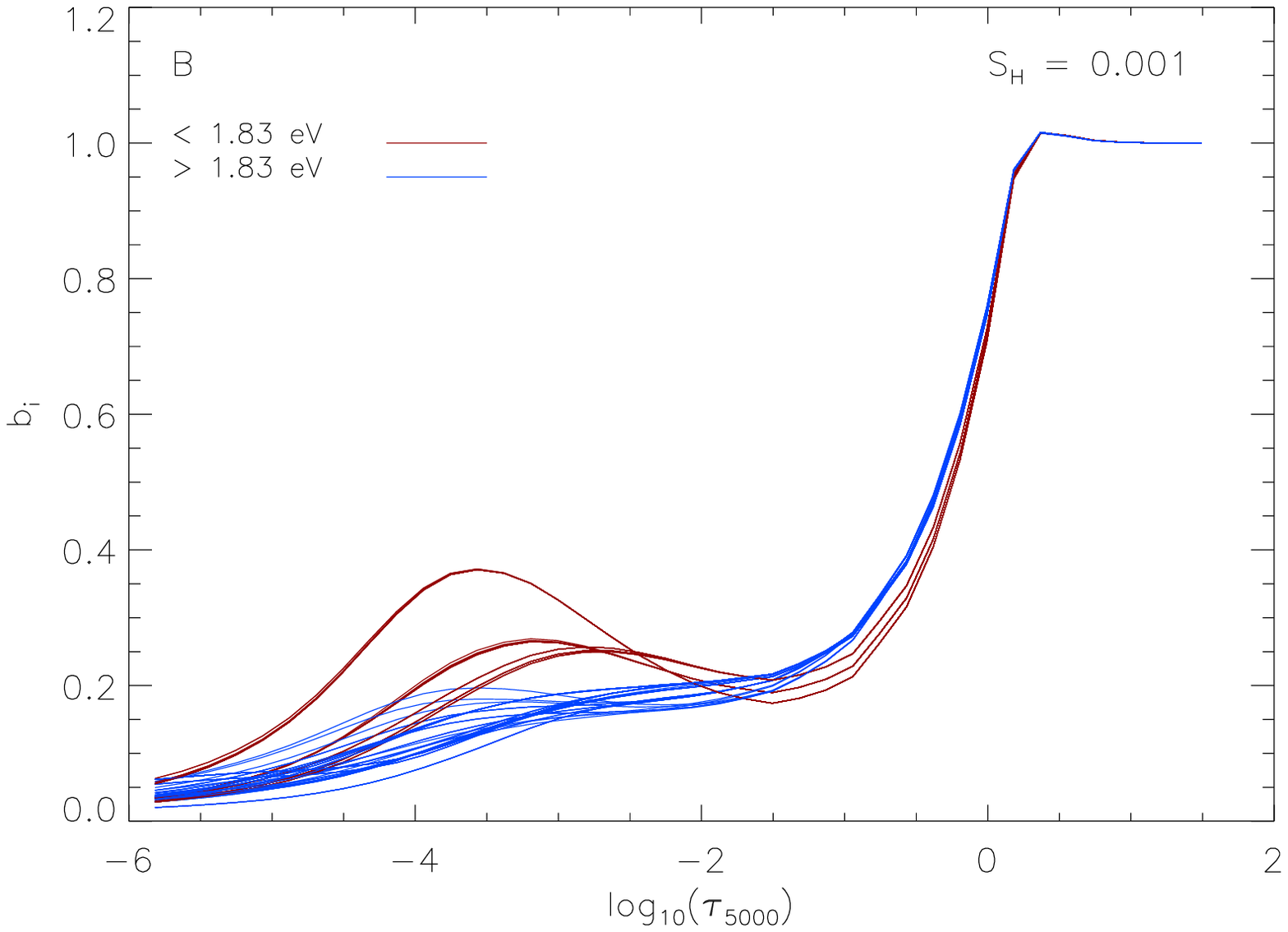,width=9cm,height=5.5cm}
	\psfig{file=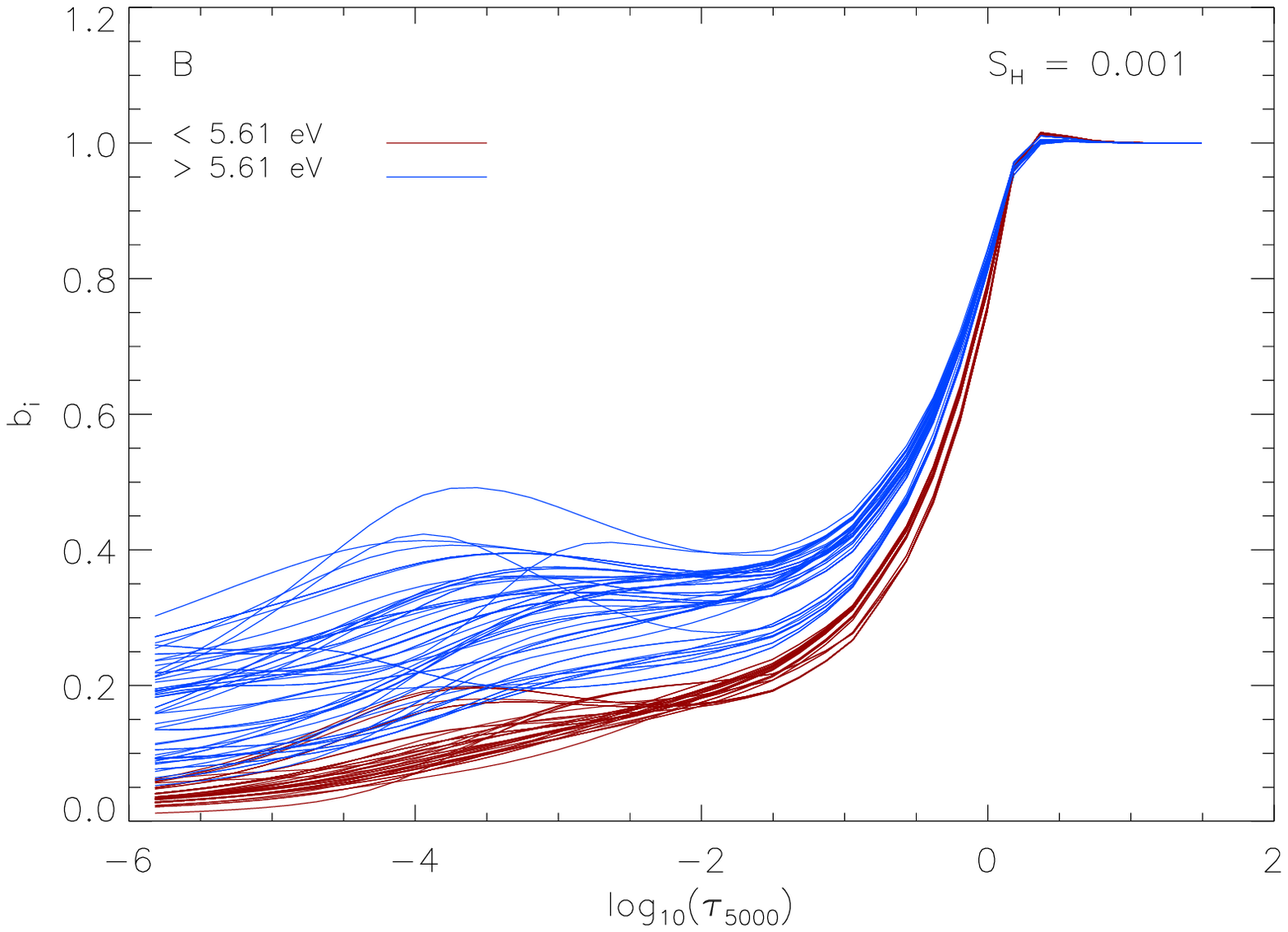,width=9cm,height=5.5cm}}
	}
\vbox{
	\hbox{
	\psfig{file=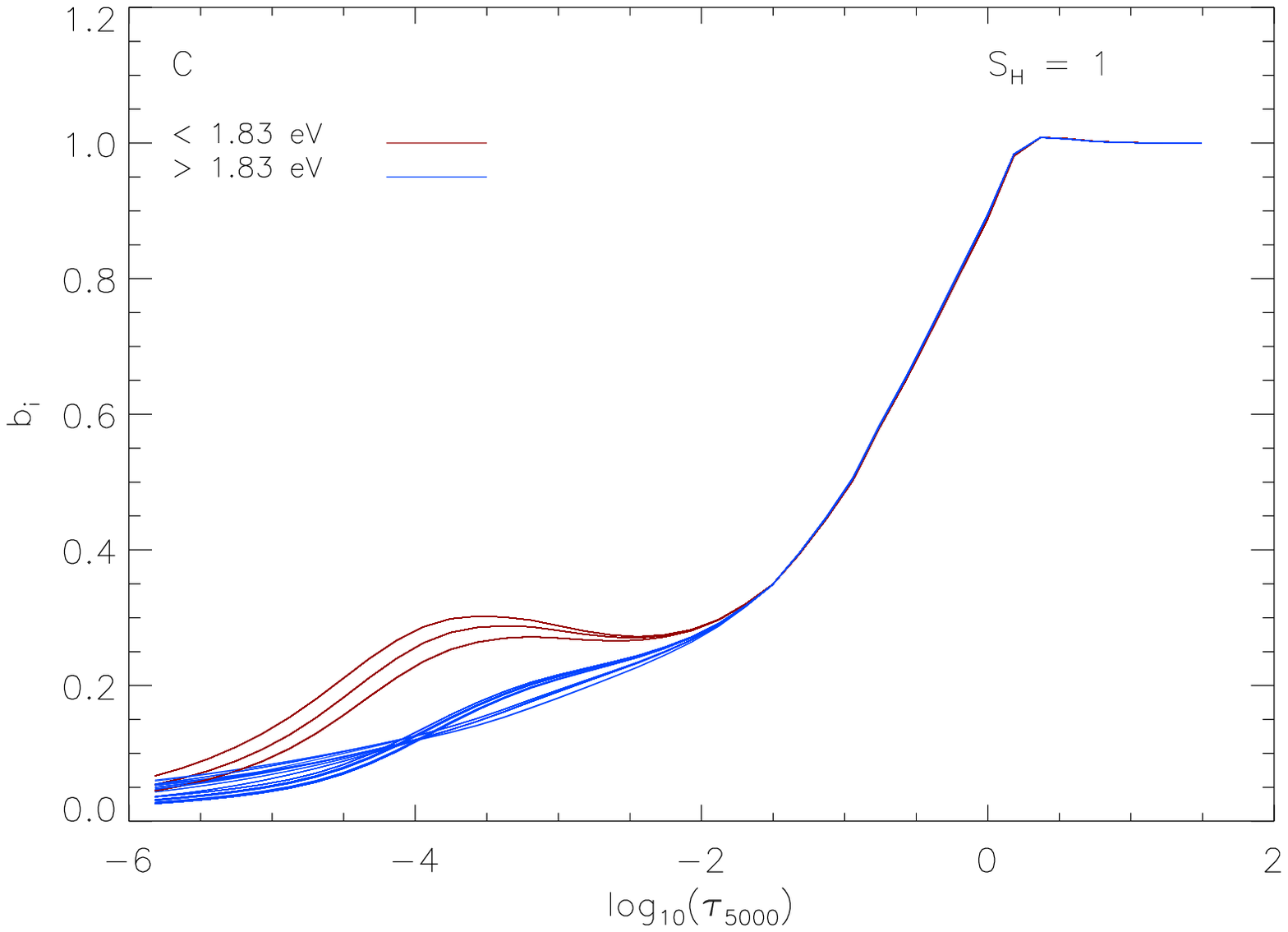,width=9cm,height=5.5cm}
	\psfig{file=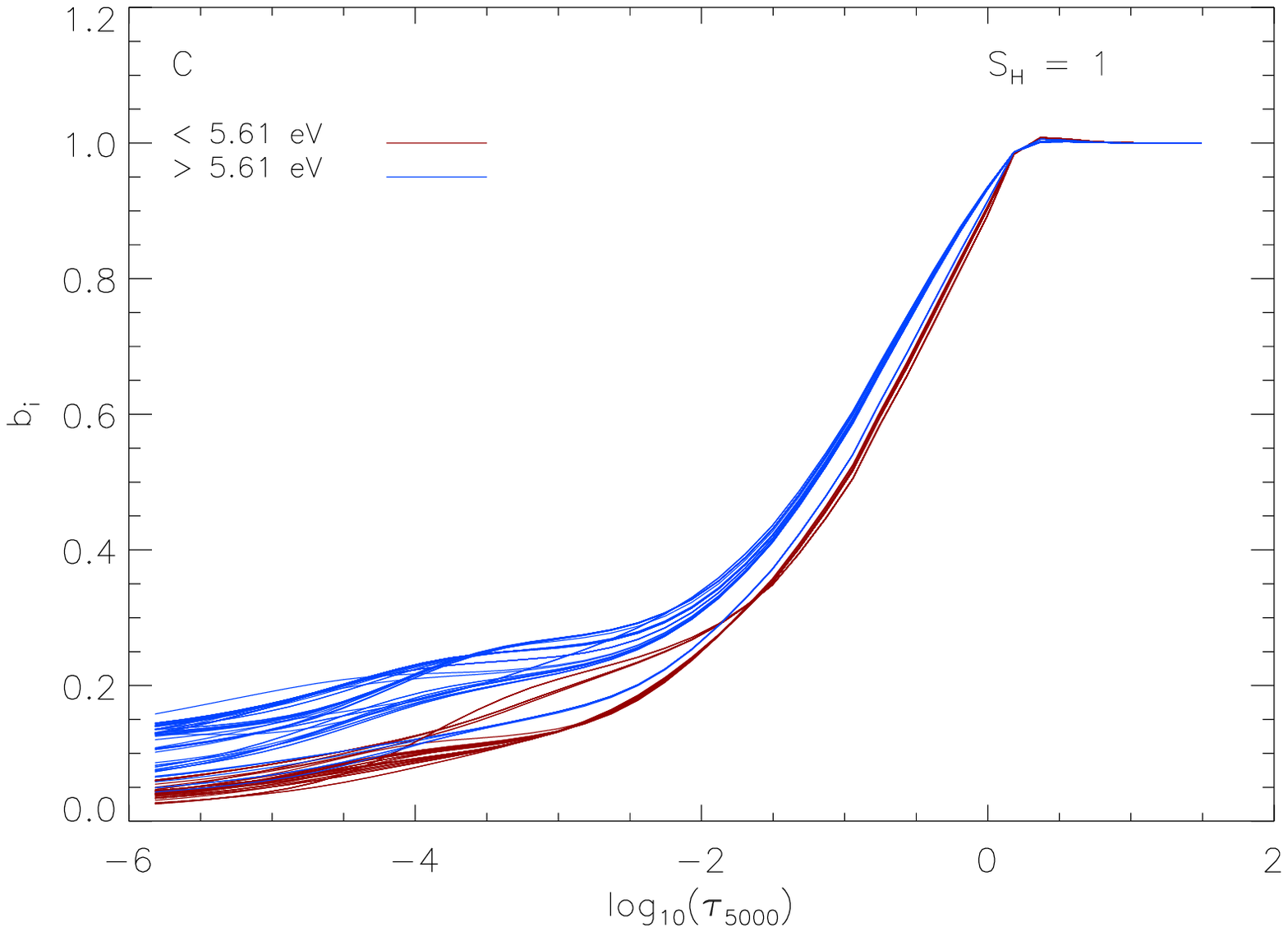,width=9cm,height=5.5cm}}
	}
	\caption{Departure coefficients for all lower levels (left) and all upper levels (right) of the lines we have studied in the star HD140283. A: $\rm S_{H}$ = 0, B: $\rm S_{H}$ = 0.001, C: $\rm S_{H}$ = 1}
	\label{fig:departplots}
\end{figure*}

\begin{figure*}[h]
\centering
\vbox{
	\hbox{
	\psfig{file=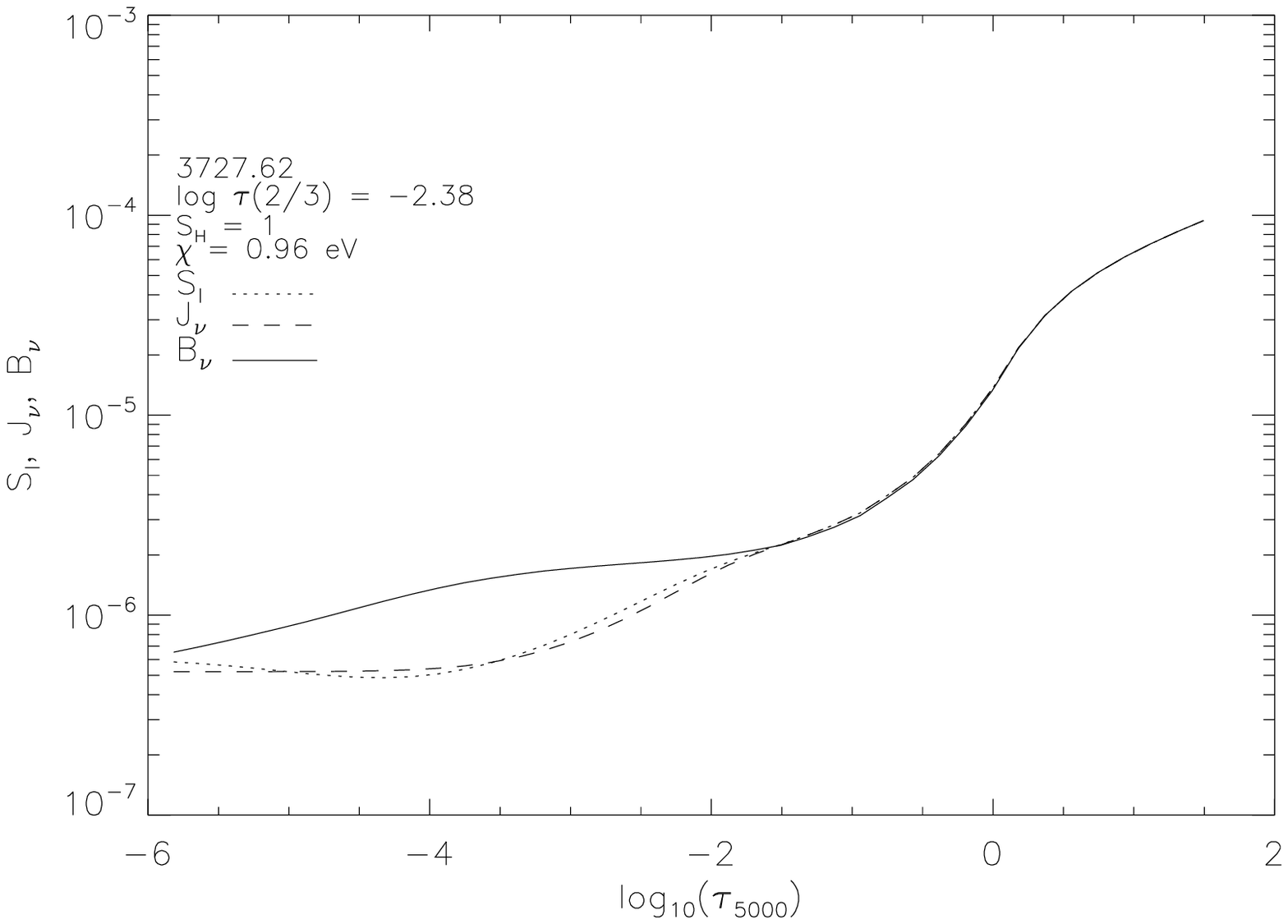,width=9cm,height=5.5cm}
	\psfig{file=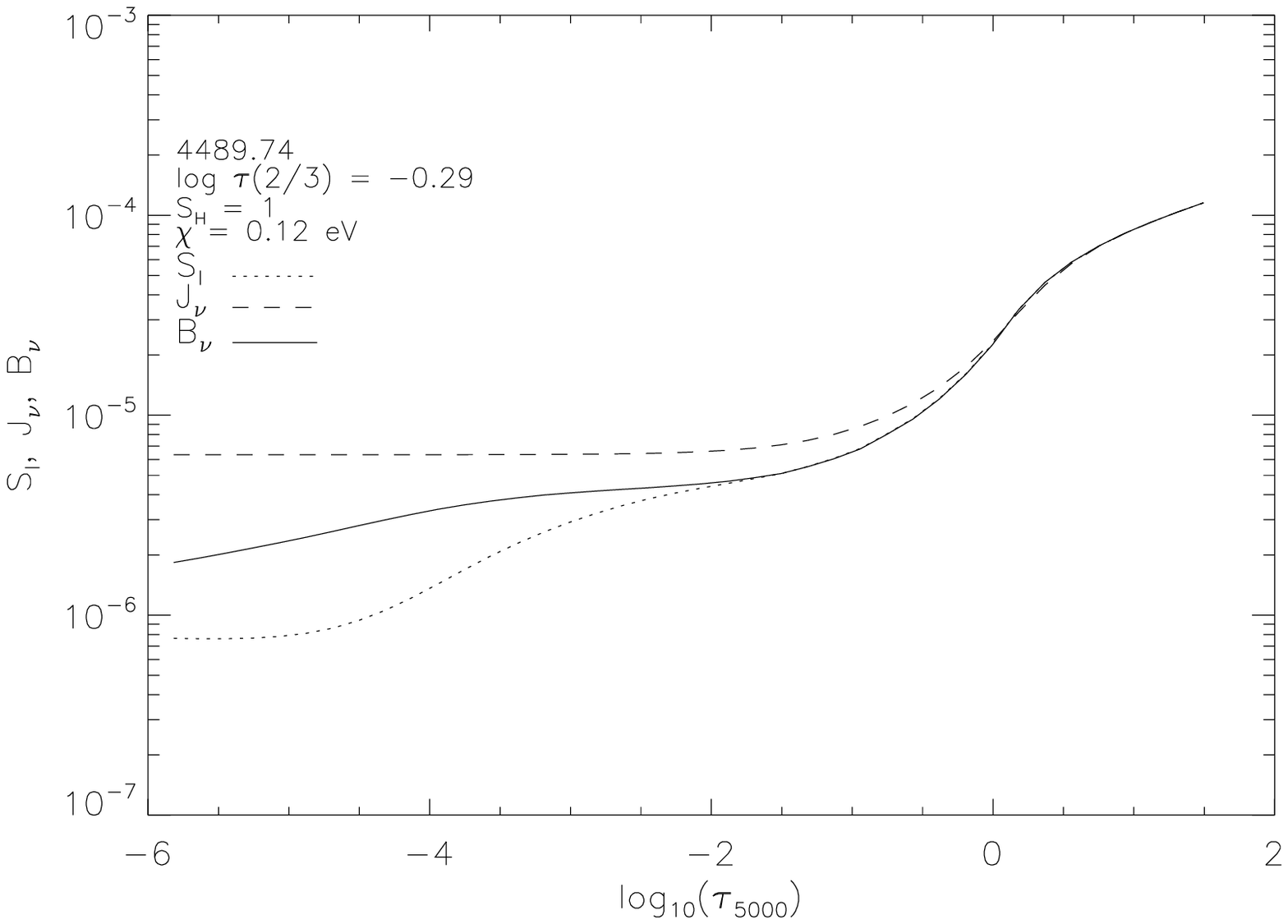,width=9cm,height=5.5cm}}
	}
	\caption{Source function,$S_{l}$, mean intensity,$J_{\nu}$, and Planck function, $B_{\nu}$, for two lines whose lower level is close to the ground state - $\chi$ = 0.96 eV (left panel) and 0.12 eV (right panel), for $\rm S_{H}$ = 1, labelled with the lines characteristic formation depth log $\tau_{5000}$ at which $\tau_{\nu}$ = 2/3.}
	\label{fig:sjbplots}
\end{figure*}

In Fig. \ref{fig:departplots}, we present the departure coefficients, $b_{i}=n_{i}/n_{i}^{\rm LTE}$, for the lower (left hand side) and upper (right hand side) levels of all lines we have measured in the star HD140283 in paper I, calculated for three $\rm S_{H}$ values. The two sets of lines in each plot, coloured. red and blue, represent levels that fall above and below the midpoint of our excitation energy range, i.e. 1.83 eV where our highest lower level of the transition is at 3.65 eV, and 5.61 eV where our highest upper transition level is at 6.87 eV. This is done to better visualise the effects of NLTE on different levels of the atom. We see that in all cases the \ion{Fe}{i} levels are under-populated compared to LTE at $\tau_{\rm 5000} < $1. This is primarily due to the effects of overionization where $J_{\nu} > B_{\nu}$ for lines formed from the levels of the atom at around $\chi \sim$ 4 eV below the continuum, due to the UV photons having energies $\approx$ 3-4 eV. This causes all levels of the atom to become greatly depopulated, as can be seen from the blue lines. The coupling of the higher levels through collisions and of the lower levels through the large number of strong lines sharing upper levels implies that relative to one another the \ion{Fe}{i} level populations approximately follow the Boltzmann distribution.  Because of photoionization, the Saha equilibrium between \ion{Fe}{i} and \ion{Fe}{ii} is not fulfilled however and the departure coefficients of \ion{Fe}{i} levels are less than unity. In deeper levels of the atmosphere, this leads to both upper and lower levels of a transition being equally affected by the above phenomena (Fig \ref{fig:departplots} -- right hand side). For this reason, the source functions for lines forming at these depths are relatively unaffected in this region, as  $S \approx (b_{\rm upper}/b_{\rm lower})B_{\nu}$, and follow a Planckian form (Fig. \ref{fig:sjbplots} -- right hand panel). The combined effect of the above processes, i.e. depopulation and relatively unaffected source functions, leads to a smaller $W_{\rm \lambda}$ and thus weaker lines, and increased abundances compared to the LTE case. For stronger lines, forming further out in the atmosphere, there is a divergence between $b_{\rm upper}$ and $b_{\rm lower}$ and the source function thus diverges from the Planck function (Fig. \ref{fig:sjbplots} -- left hand panel). In the case where $S^l_\nu < B_\nu$, the source function compensates slightly for the loss of opacity leading to smaller NLTE corrections, the opposite being true for $S^l_\nu > B_\nu$. We see that for the lower level of the weaker line considered in the figure has $\bar J_{\nu} > B_{\nu}$, whilst $S^l_\nu \approx B_\nu$, which leads to overionization of that level and greater departures than the stronger line and greater NLTE abundance corrections.

The effect of H collisions is in general to reduce the spread of departure coefficients and drive populations towards LTE values. This reduction in the spread of departure coefficients comes from the coupling of bound states. The increase of H collisions gradually reduces the departures from LTE through the atmosphere as shown in Fig. \ref{fig:departplots}; with an increasing $\rm S_{H}$ the slope in the departure coefficient profile becomes shallower. In Fig. \ref{fig:departplots} it is interesting to see that the rise in $b_{i}$ at around $\tau_{\rm 5000} \approx -2.5$ for the levels below 1.83 eV becomes smaller with increasing $S_{\rm H}$. This could in fact mean an increase in NLTE departures for some levels for increasing $S_{\rm H}$, rather than H collidions driving conditions towards LTE which is normally the case. This rise is most likely caused by increased recombination in the upper (infrared) levels followed by a cascade of electrons down to lower levels. Exactly how this is affected by the increase in $\rm S_{H}$ is not yet known and requires further study.

\begin{figure}[tbp]
\centering
	\psfig{file=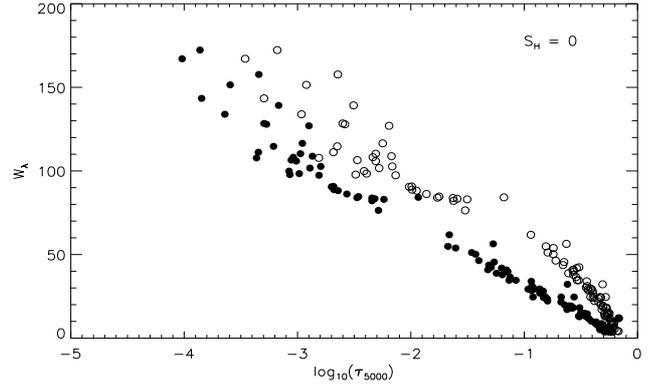,width=9cm,height=5.5cm}
	\caption{The depth of formation of \ion{Fe}{i} lines, with no H collisions, on the log $\tau_{\rm 5000}$ scale for lines of different equivalent width, in LTE (filled circles) and NLTE (open circles), for HD140283.}
	\label{fig:depthform}
\end{figure}

\begin{figure}[htbp]
\centering
	\psfig{file=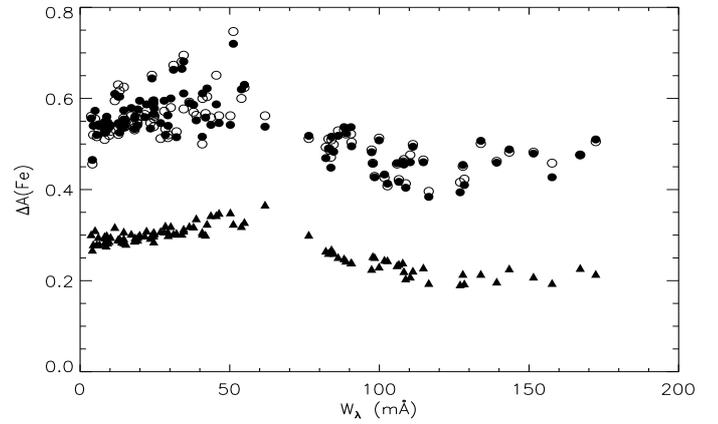,width=9cm,height=5.5cm}
	\caption{Abundance correction versus equivalent width for the lines measured in the star HD140283 for $\rm S_{H}$ = 1 (filled triangles), 0.001 (filled circles) and 0 (open circles).} 
	\label{fig:Chi-WEQHD140283}
\end{figure}

The decrease in level population at $\tau_{\rm 5000} <$ 1 causes a drop in opacity for all lines. As a result of this, the lines form deeper in the atmosphere than in LTE. In Fig. \ref{fig:depthform}, we clearly see this effect, where we show the continuum optical depth $\tau_{\rm 5000}$ at which the line optical depth $\tau_{\nu}$ = 2/3. We also see that there is an increasingly large logarithmic optical depth difference, $\Delta\log{\tau_{\mathrm{5000}}(\tau_\nu=2/3)}$, between the formation of weak lines in NLTE and LTE, up to $\approx$ 50 m\AA, after which the difference becomes constant.  With a decrease in opacity compared to LTE, there needs to be an increase of abundance to match the equivalent width of a given line in NLTE. Opacity is not the only variable affected by NLTE, the source function can also be affected. However, it is the dominant force in driving the NLTE departures within the Fe atom.  In Fig. \ref{fig:Chi-WEQHD140283}, we plot the abundance correction versus equivalent width for the star HD140283. We see that there is a positive correction for the different values of $\rm S_{H}$. There is a clear trend with equivalent width. It is how this translates to trends with excitation energy $\chi$ that will affect $T_{\rm eff}$: if the abundance corrections only shifted the mean abundance without depending on $\chi$ then the derived $T_{\rm eff}$ would not change. 

Through Fig. \ref{fig:departplots} to Fig. \ref{fig:Chi-WEQHD140283} the general effects of NLTE on line formation can be seen. The depletion of level populations (Fig. \ref{fig:departplots}) leads to a lower opacity and shifts the depth of formation to deeper levels (Fig. \ref{fig:depthform}). This also means that a higher abundance is needed within NLTE, leading to positive abundance corrections (Fig. \ref{fig:Chi-WEQHD140283}). However, there is a competing effect in some cases where the source function deviates from the Planck function (Fig. \ref{fig:sjbplots}), which, in the case of the strong lines, compensates for the level depletion and decreases the abundance correction, as is seen in Fig. \ref{fig:Chi-WEQHD140283}.

\section{NLTE abundance corrections - deriving, testing, applying}
\label{sec:NLTEAbundanceCorrectionsDerivingTestingApplying}

\begin{figure}[h]
\centering
	\psfig{file=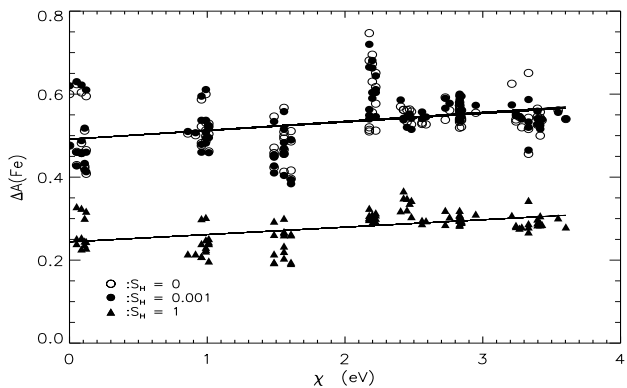,width=9cm,height=5.5cm}
	\caption{Abundance correction versus $\chi$ for $\rm S_{H}$ = 1 (filled triangles), 0.001 (filled circles) and 0 (open circles) for the measured lines in the star HD140283. Least-squares fits are shown to guide the eye.}
	\label{Fig:abnd-chi/ew-HD140283}
\end{figure}

In order to determine a new $T_{\rm eff}$ for a star, we first need to calculate NLTE corrections for the LTE abundances derived in Paper I. Abundance corrections of the form $A_{\rm NLTE,{\sc MULTI}}$ $-$ $A_{\rm LTE,{\sc MULTI}}$ are calculated and applied to the LTE abundances from Paper I to generate NLTE abundances on the same scale as that paper, rather than using solely the new NLTE analysis. This procedure is used so as to tie this work to the previous results, thus allowing the limitations of the LTE assumptions in that work to be seen. To do this, a grid of {\sc MULTI} results for a range of abundances is created with increments of 0.02 dex. The abundance values covered by this grid depend on the spread of abundances from individual lines in each star. {\sc MULTI} gives an LTE and NLTE equivalent width for each abundance in this grid. A first step is to determine what $W_{\rm LTE}$ from the {\sc MULTI} grid corresponds to the LTE abundance derived in Paper I \citep{Hosfordetal2009}. This is done for all Fe lines that are measured in the star. The NLTE abundance inferred for a line is the abundance that corresponds to this $W_{\lambda}$ within the grid of {\sc MULTI} NLTE results. The correction is then calculated as $\Delta A(\rm Fe)$ = $A(\rm Fe)_{NLTE}$ $-$ $A(\rm Fe)_{LTE}$. Fig. \ref{Fig:abnd-chi/ew-HD140283} shows the corrections for the star HD140283 calculated for the three different $\rm S_{H}$ values: $\rm S_{H}$ = 0, 0.001 and 1. We see a trend in the abundance correction with $\chi$, where we have values, from least square fits, of:
\begin{equation}
	A(\rm Fe)=0.490(\pm0.012)+0.0216(\pm0.0055)\chi\rm ;\  for\  S_{H} = 0
\end{equation}
\begin{equation}
		A(\rm Fe)=0.490(\pm0.012)+0.0207(\pm0.0054)\chi\rm ;\  for\  S_{H} = 0.001\\
\end{equation}
\begin{equation}
		A(\rm Fe)=0.244(\pm0.072)+0.0178(\pm0.0032)\chi\rm ;\  for\  S_{H} = 1\\
\end{equation}
The non-zero coefficient of $\chi$ implies that a $T_{\rm eff}$ correction is needed. The values for $\rm S_{H}$ = 0 and $\rm S_{H}$ = 0.001 are very similar and imply that $T_{\rm eff}$ corrections for these two values will be very similar. We therefore decided that corrections for only $\rm S_{H}$ = 0 and 1 would be calculated, $\rm S_{H}$ = 0 representing the maximal NLTE corrections and $\rm S_{H}$ = 1 representing the full Drawinian magnitude of neutral H collisions.

To test the corrections, we compared synthetic profiles from the NLTE abundance with the observed profile, and compared measured $W_{\lambda}$'s with NLTE $W_{\lambda}$'s from {\sc MULTI}, obtained from an abundance given by $A_{\rm LTE}$ + $\Delta A$. The synthetic profiles are convolved with a Gaussian whose width is allowed to vary from line to line. This represents the macroturbulent and instrumental broadening, the latter calculated by fitting Gaussian profiles to ThAr lines in IRAF and found to be $\sim$ 100 m\AA. We found that the profiles match the observed line reasonably well, and that measured and {\sc MULTI} calculated $W_{\lambda}$'s are comparable, with a standard deviation of 2.3 m\AA. This gives us confidence that the corrections are realistic within the framework of the atomic model used. These corrections were then applied to the WIDTH6 LTE abundances used in Paper I and new plots of $\chi$ versus $A$(Fe) were plotted. We then nulled trends in this plot to constrain $T_{\rm eff}$(NLTE) by recalculating the LTE abundances using the radiative transfer program WIDTH6 (Kurucz \& Furenlid 1978) exactly as in \citet{Hosfordetal2009} and reapplying the NLTE corrections, derived here from {\sc MULTI} for the original LTE parameters.

\begin{figure}[h]
\centering	
\vbox{
	\psfig{file=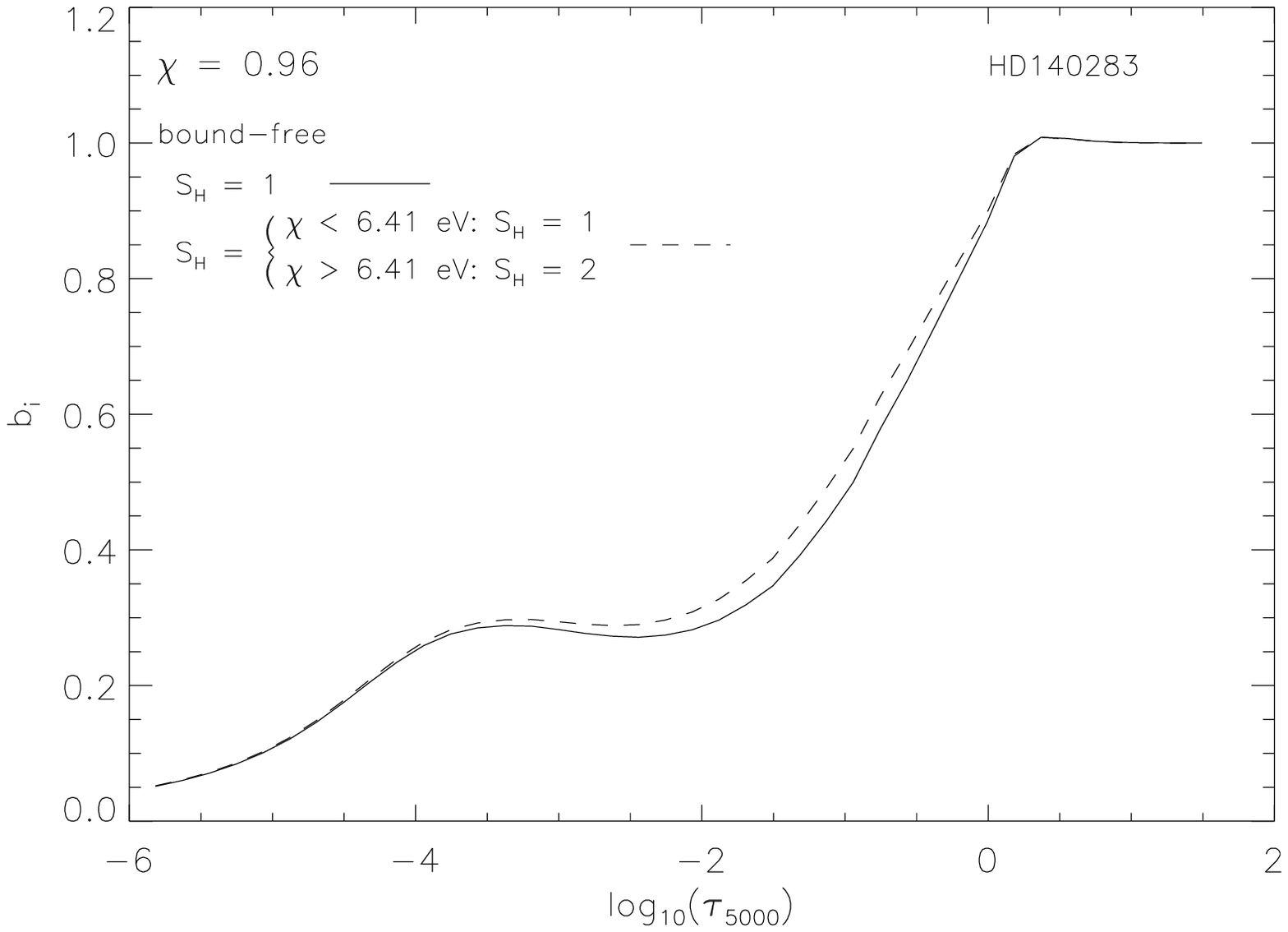,width=9cm,height=5.5cm}
	\psfig{file=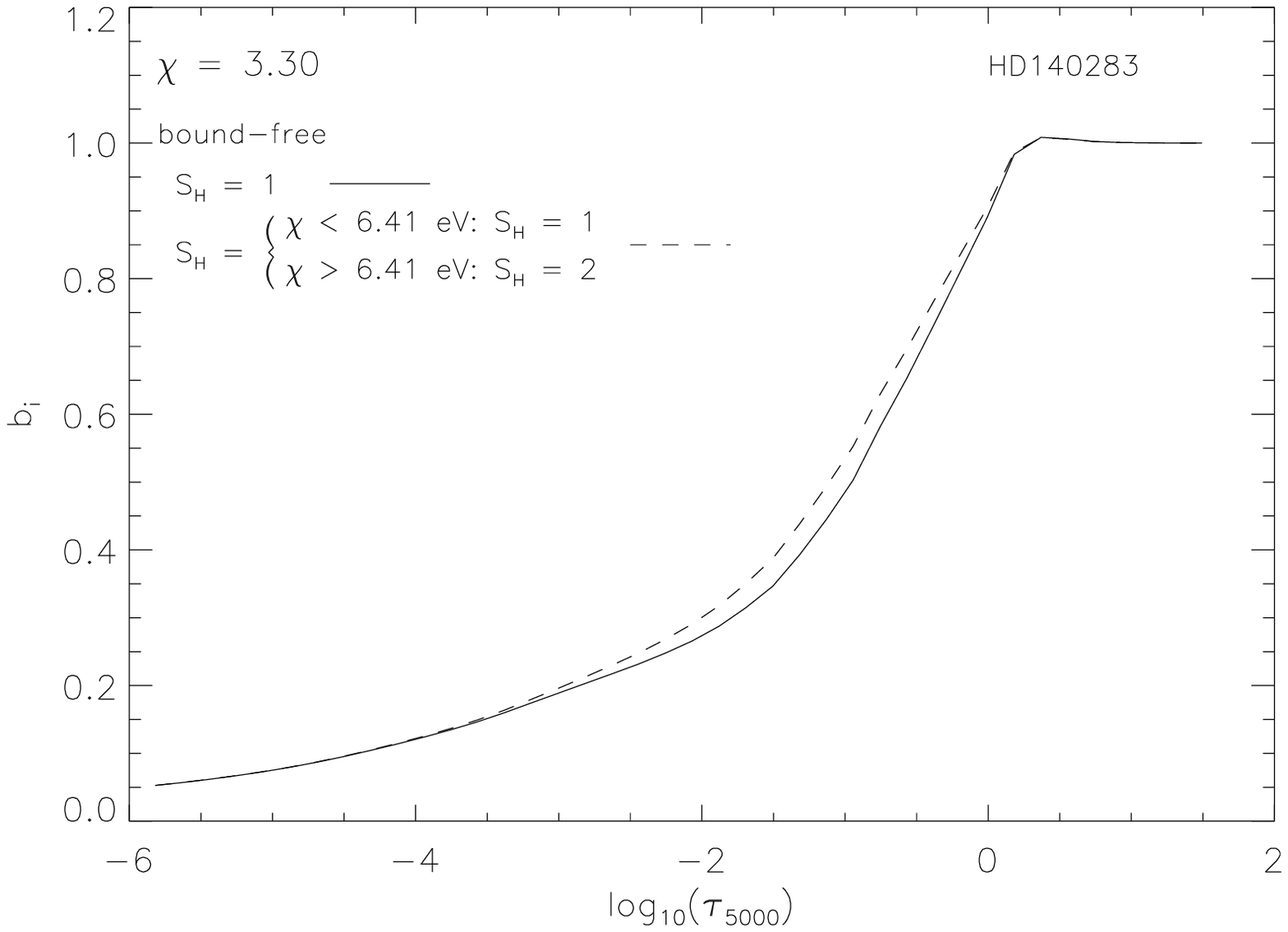,width=9cm,height=5.5cm}}
	\caption{Dashed line: The effects of increasing the $\rm S_{H}$ value to 2 for the top 0.5 eV of levels in the atomic model, for  bound-free transitions from a low level (top panel) and a higher level (bottom panel). The remaining levels have $\rm S_{H}$ = 1. Solid line: $\rm S_{H}$ = 1 for all levels.}
	\label{fig:ULeffects}
\end{figure}

\begin{figure} [h]
	\psfig{file=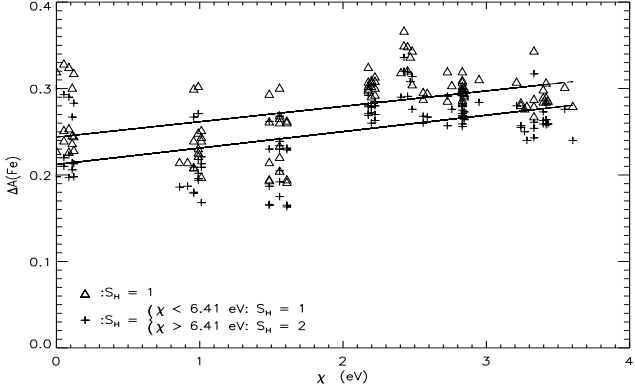,width=9cm,height=5.5cm}
	\caption{Comparison between the abundance correction versus excitation energy for the star HD140283 using $\rm S_{H}$ = 2 for the top 0.5 eV of levels of \ion{Fe}{i} in the atomic model, and $\rm S_{H}$ = 1 for the remaining levels (crosses) and pure $\rm S_{H}$ = 1 (open triangles).}
	\label{fig:chiAULeffect}
\end{figure}

As noted in Sect. \ref{sec:transitionrates}, it can be important to include the highest levels of the atom in the calculations. It is not necessary to include each individual level however, and it is possible to use superlevels that represent groups of closely spaced levels \citep{Korn2008}. To test the effect of these upper levels, we took the approach of giving the top 0.5 eV of levels in our atomic model an $\rm S_{H}$ = 2 whilst the rest of the levels had $\rm S_{H}$ = 1.  We have done this for three situations; A) increasing $\rm S_{H}$ for just the bound-bound transitions rates, B) increasing $\rm S_{H}$ for just the bound-free rates and C) increasing $\rm S_{H}$ for both the bound-bound and bound-free. We discuss here only the case of the bound-free rates as it is only these rates that have an effect, edging the populations towards LTE values. Changing the bound-free rates not only affects the higher levels but translates through all lower ones. In fact it is the lower half of the atomic model that is affected by a greater amount; further investigation into reasons for this effect are discussed in Sect. \ref{sec:TheTRmEffScale}. The result can be seen in Fig. \ref{fig:ULeffects} where we plot a level with $\chi$ = 0.96 eV and one of the higher levels, $\chi$ = 3.30 eV,  from our atomic model. Fig. \ref{fig:chiAULeffect} shows the abundance correction against $\chi$ for the increased $\rm S_{H}$ value of the upper levels and for a pure $\rm S_{H}$ = 1 situation. Comparing the differences in abundance correction between $\rm S_{H}$ = 1, and $\rm S_{H}$ = 1 with $\rm S_{H}$ = 2 on the upper levels we see a mean difference ($\Delta A(\rm Fe)_{S_H=1+2}$ -- $\Delta A(\rm Fe)_{S_{H}=1}$) of $-0.031$ dex for $\chi$ = 0 - 2 eV and $-0.028$ dex for $\chi >$ 2 eV, for the star HD140283. These effects equate to a 5 K increase in $T_{\rm eff}$ compared to $\rm S_{H}$ = 1. It is then clear that the upper levels have a slight effect on the final temperatures, and induce a slightly larger NLTE correction. However, in the case of this study, where random errors are of order $\sim$ 80 K, they will not make a significant effect.

%______________________________________________________________

\section{Results}
\label{sec:Results}

Fe abundance corrections for the stars in Table \ref{Table1} have been calculated and new temperatures have been derived using the excitation energy technique, as in Paper I but with the NLTE corrections applied as described in Sect. \ref{sec:NLTEAbundanceCorrectionsDerivingTestingApplying}. Table \ref{Table2} lists the new NLTE $T_{\rm eff}$'s and $\Delta T_{\rm eff}$, such that $\Delta T_{\rm eff}$ = $T_{\rm eff}(NLTE)$ $-$ $T_{\rm eff}(LTE)$, for the selection of stars.
\begin{table*}[tbp]
\caption{Final $T_{\rm eff}$ and $A$(Li) for the selection of stars in this study}
\label{Table2}
	\centering
		\begin{tabular} {l c c c c c c c c c}
			\hline\hline
			&  &  &\multicolumn{3}{c}{(NLTE) $S_{H}$ = 0} & & \multicolumn{3}{c}{(NLTE) $S_{H}$ = 1} \\
			\cline{4-6}  \cline{8-10}
			Star	&	$T_{\rm eff}$  (LTE)	& log $g$  &	$T_{\rm eff}$ 	&	 $\Delta T _{\rm eff}$ 	&	$A$(Li)  &	&	$T_{\rm eff}$  &	$\Delta T_{\rm eff}$ 	&	$A$(Li) 	\\
						&		(K)		&    (dex)   &  (K)   &  (K)  &  (dex)   &  &(K)  &  (K)  & (dex) \\
			\hline
			HD140283	&	5769	&	3.73 & 5850 $\pm$ 51	&	81	&	2.27 $\pm$ 0.03 &	&	5838 $\pm$ 48	&	69	&	2.26 $\pm$ 0.03	\\
			HD84937	&	6168	&	3.98 &6318 $\pm$	102 &	150	&	2.28 $\pm$ 0.07 &	&	6261 $\pm$ 102	&	93	&	2.24 $\pm$ 0.07	\\
			HD74000	&	6070 & 4.03	&6211 $\pm$ 131	&	141	&	2.15 $\pm$ 0.09 &	&	6145 $\pm$ 130	&	75	&	2.10 $\pm$ 0.09	\\
			BD$-$26$^{\circ}$2621	&	6233 & 4.49	&	6298 $\pm$ 81	&	65	&	2.21 $\pm$ 0.05 &	&	6292 $\pm$ 80	&	59	&	2.21 $\pm$ 0.05	\\
			CD$-$33$^{\circ}$1173	&	6386& 4.44	&	6293 $\pm$ 128	&	$-$93	&	2.09 $\pm$ 0.08 &	&	6427 $\pm$ 113	&	41	&	2.18 $\pm$ 0.07	\\
			LP815-43 (SGB)	&	6400 & 3.87 &	6402 $\pm$ 100	&	2	&	2.13 $\pm$ 0.07 &	&	6522 $\pm$ 119	&	122	&	2.21 $\pm$ 0.08	\\
			LP815-43 (MS)	&	6529 & 4.40	&	6551 $\pm$ 102	&	22	&	2.23 $\pm$ 0.07 &	&	6607 $\pm$ 99	&	78	&	2.27 $\pm$ 0.06	\\
			\hline
		\end{tabular}
\end{table*}

For this work, all the other parameters, viz. log $g$, [Fe/H] and $\xi$, were kept at the values found in \citet{Hosfordetal2009}. Our aim here, as it was in \citet{Hosfordetal2009}, is to narrow down the zero point of the temperature scale by quantifying the systematic errors, albeit at the expense of having larger star to star random errors. Contributions to the errors come from adopted gravity, the nulling procedure in determining the $T_{\rm eff}$,  and smaller contributions from the error in microturbulence, errors in the age, metallicity and initial temperature, $T_{\rm phot}$, when determining isochronal gravities. In relation to the gravities, the three HD stars had gravities derived using {\sc hipparcos} parallaxes, and their errors are a reflection of errors propagating through this calculation, whilst for the remaining stars isochrones were used. The isochronal gravities are sensitive to age, with a 1 Gyr difference leading to a change of $\sim$ 0.03 dex for main sequence (MS) stars and $\sim$ 0.06 dex for sub-giant (SGB) stars. This equates to a change in $T_{\rm \chi}$ of 12 K and 24 K respectively. These errors are based on LTE sensitivities, as are other errors quoted below. There is also a dependence on the initial temperature, a photometric temperature from Ryan et al (1999), used to determine the isochronal gravity. A +100 K difference leads to +0.06 dex and --0.06 dex for MS and SGB stars respectively. This equates to $\pm$ 24 K in $T_{\rm \chi}$ which shows, importantly, that $T_{\rm \chi}$ is only weakly dependent on the initial photometric temperature. Contributions to $T_{\rm \chi}$ is also sensitive to microturbulence, for which an error of $\sim$ 0.1 km $\rm s^{-1}$ equates to an error of $\approx$ 60 K.

In the nulling procedure any trends between [Fe/H] and $\chi$ are removed. Due to the range in line to line Fe abundances for a particular star, there is a statistical error in the trend which is of order $\sigma$ = 0.011 dex per eV, which equates to $\approx$ 40 K - 100 K depending on the star under study. This error also contains the random line-to-line errors due to equivalent width, $gf$, and damping values. The final $T_{\rm eff}$ error in Table \ref{Table2} is then a conflation of this statistical error and the errors from $\Delta$age = 1 Gyr, $\Delta\xi$ = 0.1 km$\rm s^{-1}$, $\Delta$[Fe/H] = 0.05 and $\Delta T_{\rm phot}$ = 100 K.

These new $T_{\rm eff}$ values and equivalent widths from \citet{Ryanetal1999} were then used to calculate new \element[][]{Li} by interpolating within a grid of equivalent width versus abundance for different $T_{\rm eff}$. This grid was taken from \citet{Ryanetal1996a}.

\section{Discussion}
\label{sec:Discussion}

\subsection{The $T_{\rm eff}$ scale}
\label{sec:TheTRmEffScale}

\begin{figure}
\centering
\vbox{
		\psfig{file=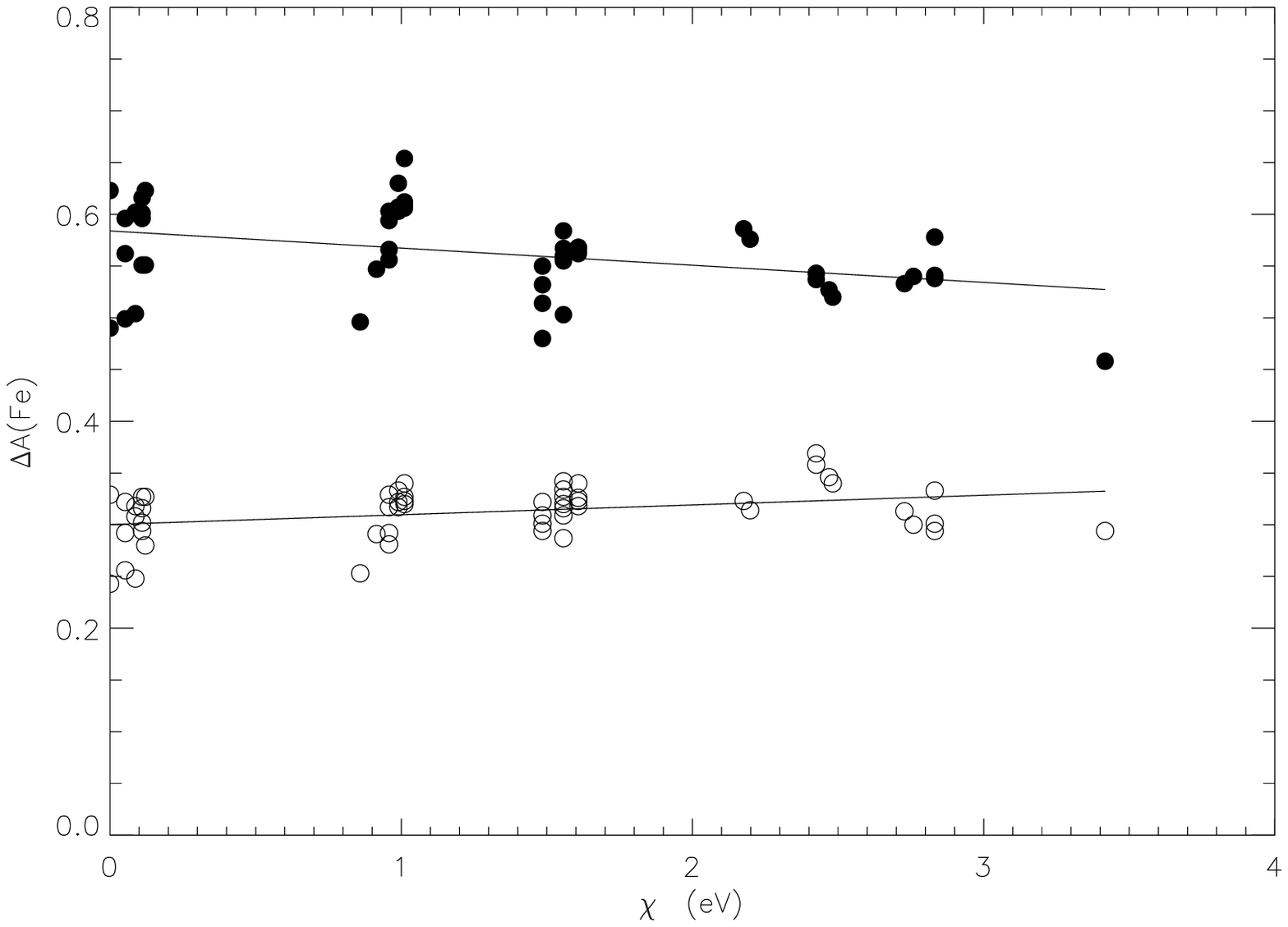,width=9cm,height=5.5cm}
		\psfig{file=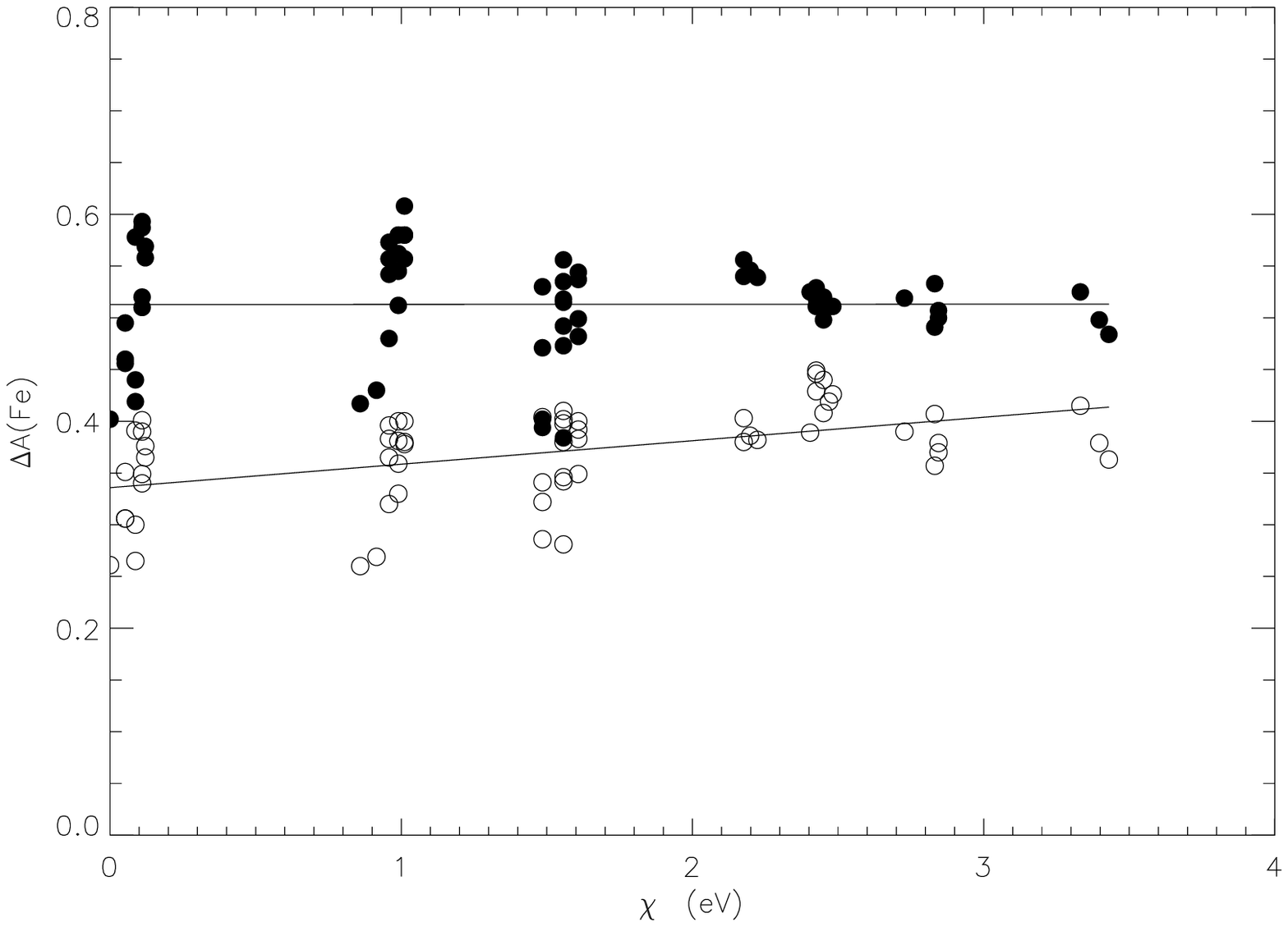,width=9cm,height=5.5cm}}
		\caption{Abudance correction versus $\chi$ for (top panel) CD$-$33$^{\circ}$1173 and (bottom panel) LP815$-$43 for $\rm S_{H}$ = 0 (filled circles) and $\rm S_{H}$ = 1 (open circles).}
		\label{Fig:CD/LP-chiAbund}
\end{figure}

With the addition of the NLTE corrections, we see in Table \ref{Table2} that there is, for the most part, an increase in $T_{\rm eff}$ from the LTE $T_{\rm eff}$'s of \citet{Hosfordetal2009}, for both cases of $\rm S_{H}$. The only exception is CD$-$33$^{\circ}$1173 in the $\rm S_{H}$ = 0 case, for which there is a 93 K decrease. We return to this star below. The $T_{\rm eff}$ corrections we have derived average 59 K for $\rm S_{H}$ = 0 and 73 K for $\rm S_{H}$ = 1 (treating LP815$-$43 as one datum, not two). For $\rm S_{H}$ = 0 the temperature corrections tend to increase at cooler temperatures, whilst the tendency is weaker or opposite for $\rm S_{H}$ =1, i.e. corrections increase at higher temperatures (obviously the gravity and metallicity of the stars also affects their NLTE corrections, but nevertheless we find it intsructive to consider temperature as one useful discriminating variable). This gives rise to a change in the difference $\Delta T_{\rm eff, S_{H} = 0} - \Delta T_{\rm eff, S_{H} = 1}$ with temperature, with this quantity being negative for the two hottest stars, CD$-$33$^{\circ}$1173 and LP815$-$43. The switch over from $\rm S_{H}$ = 0 having the larger correction to $\rm S_{H}$ = 1 having the larger correction is at around $T_{\rm eff} \approx$ 6200 K . Further testing has shown that this is not a random error and is clearly something to investigate further in the future. This is further shown by Fig. \ref{Fig:CD/LP-chiAbund} where the abundance correction versus $\chi$ for the stars CD$-$33$^{\circ}$1173 and LP815$-$43 (SGB) are plotted. It is seen that for LP815$-$43 (SGB), increasing $\rm S_{H}$ has a larger effect on the lower excitation lines than for higher ones. This has induced a trend of abundance with $\chi$ larger than that of the $\rm S_{H}$ = 0 case. This in turn leads to a larger temperature correction for $\rm S_{H}$ = 1 than for $\rm S_{H}$ = 0. The reason for this effect is still uncertain. 

To investigate this behaviour further, the test of increasing the $\rm S_{H}$ value of the upper levels, as done on HD140283 in Sect 4., has also been performed on LP815$-$43 for the MS and SGB parameters. This has shown that the effect of collisions with neutral H are indeed larger for the lower levels of the atom, and that this effect is larger for LP815$-$43 (MS), which is the hottest star. This indicates that there is a temperature dependence, i.e. the difference between the mean difference ($\Delta A(\rm Fe)_{S_H=1+2}$ -- $\Delta A(\rm Fe)_{S_{H}=1}$) (where $\rm S_{H}=1+2$ indicates the scenario of having $\rm S_{H} = 2$ for the top 0.5 eV worth of levels) for the levels with $\chi < 2$ eV and those with $\chi > 2$ eV is greater for the hotter star, LP815$-$43 (MS). However, when performing this test on LP815$-$43 (SGB), which has a similar log $g$ to HD140283 whilst still being hotter, the effect is not as great as for HD140283. This shows that there is some gravity dependence on the neutral H collisions along with the temperature dependence i.e. the gravity indirectly affects the collisional rates, by impacting on the number density of hydrogen atoms at a given optical depth. Fig. \ref{Fig:CD/LP-chiAbund}, along with Fig. \ref{Fig:abnd-chi/ew-HD140283}, clearly show that NLTE has varying star to star effects, i.e. from the similar effects at different $\rm S_{H}$ values in HD140283 (Fig. \ref{Fig:abnd-chi/ew-HD140283}), to the differing effects in CD$-$33$^{\circ}$1173 and LP815$-$43 (SGB) (Fig. \ref{Fig:CD/LP-chiAbund}). The range of $\Delta T_{\rm eff}$ values, and the negative value for CD$-$33$^{\circ}$1173, shows the intricacies of the NLTE process, and that generalisations are not easily made when identifying the effects of NLTE on temperatures determined by the excitation energy method. For the purposes of this paper, which is concerned with the effective temperatures in the context of the available NLTE model, it is appropriate to acknowledge these NLTE effects and to move ahead to use them in the study of the Li problem, whilst still recognising that much work remains before we approach a complete description of the Fe atom.

Although we discussed the possibility that the extreme (negative) $\Delta T_{\rm eff}$ correction for CD$-$33$^{\circ}$1173 is due to corrrections being temperature-dependent, this unusual case may be in part due to the fact that only a subset of the original lines measured is available through the NLTE atomic model. The atomic model does not contain every level of the Fe atom and therefore some transitions are not present in the calculations. This means that not every line measured for a given star is present in the calculations and leads to a trend being introduced in the $\chi$-abundance plot prior to the trend induced by the NLTE corrections. This is because the original nulling of the $\chi$-abundance plot was achieved with a greater number of points. CD$-$33$^{\circ}$1173 has the least lines available from the atomic model used with {\sc MULTI}, however, there is no distinct trend between $\Delta T_{\rm eff}$ and the number of lines available for each star, and after testing we found that the effect of the subset, i.e. the measured lines that are available with our atomic model, is to increase the LTE temperature. This implies that the decrease in $T_{\rm eff}$ for this star is most likely due to NLTE effects. Although there is no obvious correlation between the number of lines available and the temperature correction, this emphasises the need for a complete atomic model. This is especially true when considering the abundance of individual lines, as in the excitation technique used in this work.
 
\begin{figure}[tbp]
\centering
\vbox{
	\psfig{file=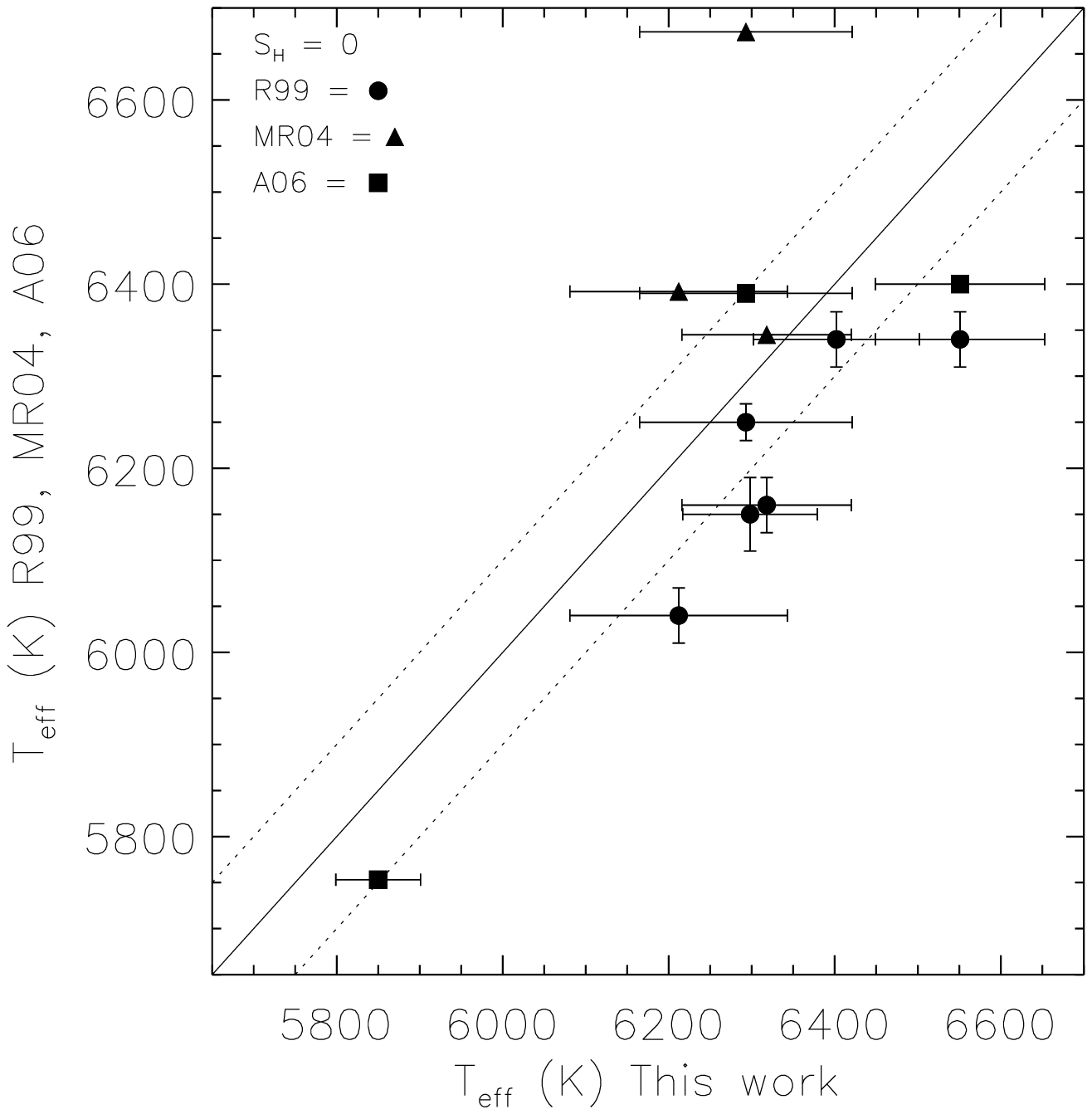,width=9cm,height=8cm}
	\psfig{file=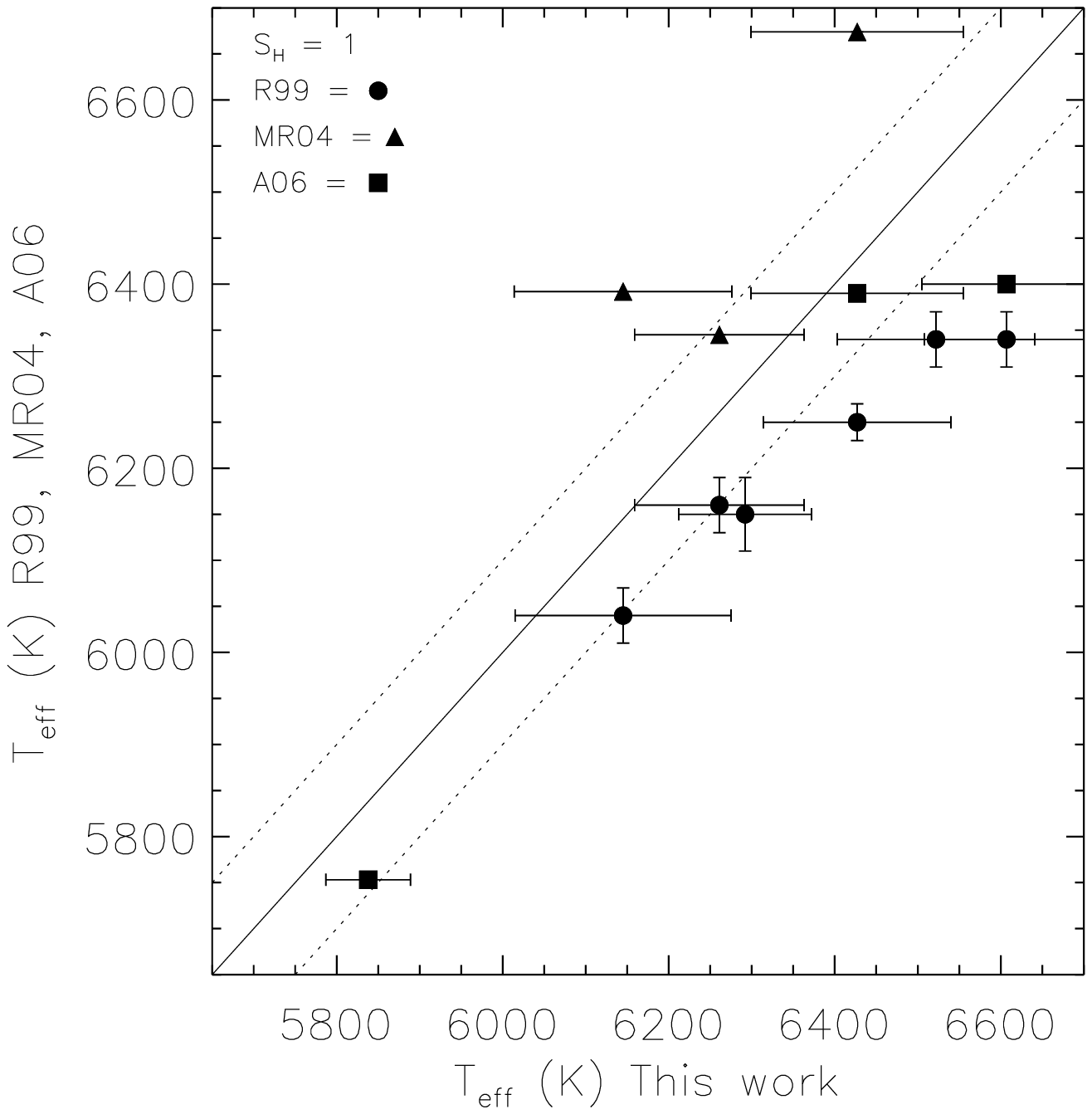,width=9cm,height=8cm}}
	\caption{$T_{\rm eff}$ comparison between this work and \citet[R99]{Ryanetal1999} (filled circles), \citet[MR04]{MelendezRamirez2004} (filled triangles) and \citet[A06]{Asplundetal2006} (filled squares), for (top panel) $\rm S_{H}$ = 0 and (bottom panel) $\rm S_{H}$ = 1. Dashed lines are $\pm$ 100 K limits.}
	\label{Fig:TempComp}
\end{figure}

As in Paper I, we have compared our $T_{\rm eff}$ values with those of \citet{Ryanetal1999}, \citet{MelendezRamirez2004}, and \citet{Asplundetal2006}. Fig. \ref{Fig:TempComp} presents these comparisons. Comparing against the photometric temperatures of \citet{Ryanetal1999} for five stars in common, we see that our new $T_{\rm eff}$ scale is hotter by an average of 132 K, with a minimum and maximum of 43 K and 211 K respectively for an $\rm S_{H}$ = 0. Recall that $\rm S_{H}$ = 0 corresponds to the maximal NLTE effect, i.e. no collisions with the hydrogen, for the model atom we have adopted. For $\rm S_{H}$ = 1, our scale is hotter by an average of 162 K, with a minimum and maximum of 101 K and 267 K respectively.

We have three stars in common with \citet{MelendezRamirez2004}. Their temperatures are hotter than the ones we derived here by 196 K on average for $\rm S_{H}$ = 0 with a minimum and maximum difference of 27 K and 381 K respectively, and by 193 K on average for $\rm S_{H}$ = 1, with a minimum and maximum difference of 84 K and 247 K respectively. Therefore, even with NLTE corrections we still cannot achieve the high $T_{\rm eff}$ of the \citet{MelendezRamirez2004} study. It has however been noted (Mel$\rm \acute{e}$ndez 2009 - private communication) that the \citet{MelendezRamirez2004} temperatures suffer from systematic errors due a imperfect calibration of the bolometric correction for the choice of photometric bands used. This led to an inaccurate zero point and hotter $T_{\rm eff}$'s than most other studies. The revision of their temperature scale is not yet available and comparisons to their new $T_{\rm eff}$'s is not possible at this time.

Finally, we have three stars in common with \citet{Asplundetal2006}. Using $\rm S_{H}$ = 0 we obtain temperatures for two of the stars that are hotter than \citet{Asplundetal2006} by 97 K and 151 K. The third star is CD$-$33$^{\circ}$1173, for which we calculated a negative temperature correction, and which is cooler in our study by 97 K. The temperatures for all three stars are hotter in our study than in \citet{Asplundetal2006} when using $\rm S_{H}$ = 1. Here the average difference is 110 K, values ranging from 37 K to 207 K. If the \citet{Asplundetal2006} temperatures are affected by NLTE, as stated by \citet{Barklem2007} who expects a 100 K increase in Balmer line temperatures, this would bring the $T_{\rm eff}$ scales back into agreement. Another problem facing the Balmer line method is the effects of granulation, due to convection, on the line wings \citep{Ludwigetal2009}. It has been found (Bonifacio -- private communication) that inclusion of these effects would increase the effective temperatures derived with this method. In particular a value of  $T_{\rm eff}$ = 6578 K has been found for the star LP815$-$43. Although this is 176 K hotter than our result for the SGB case with $\rm S_{H}$ = 0, i.e. $T_{\rm eff}$ = 6402 K, it is in good agreement with the values $T_{\rm eff}$ = 6522 K ($\rm S_{H}$ = 1) for the SGB case and $T_{\rm eff}$ = 6551 K ($\rm S_{H}$ = 0) or $T_{\rm eff}$ = 6607 K ($\rm S_{H}$ = 1) for the MS case, calculated in this work.

\subsection{Lithium abundances}
\label{sec:LithiumAbundances}

We now address the new Li abundances and their effect on the lithium problem. We see that the introduction of NLTE corrections to the $T_{\rm eff}$ scale has led to temperatures that are of order 100 K hotter than LTE temperature scales, with the obvious exception of the \citet{MelendezRamirez2004} scale. This will then lead to an increase in the mean lithium abundance. Table \ref{Table2} lists $A$(Li) for the new temperatures. With these new $T_{\rm eff}$'s, we calculate a mean Li abundance of $A$(Li) = 2.19 dex with a scatter of 0.072 dex when using $\rm S_{H}$ = 0, and $A$(Li) = 2.21 dex with a scatter of 0.058 dex for the $\rm S_{H}$ = 1 case. Consistent with the temperature increase, these values are higher than those found by other studies, in particular \citet{Spiteetal1996}, who found a value of $A$(Li) = 2.08 ($\pm$0.08) dex using a similar iron excitation energy technique but without the NLTE corrections, \citet{Bonifacioetal2007} with $A$(Li) = 2.10 ($\pm$0.09) using a Balmer line wing temperature scale, and $A$(Li) = 2.16 dex or $A$(Li) = 2.10 depending on the evolutionary state from \citet{Hosfordetal2009}. The NLTE corrections have moved the mean Li abundance closer to, but not consistent with, the WMAP value of $A$(Li) = 2.72 dex, and thus still leaves the lithium problem unsolved. It is noted that even the \citet{MelendezRamirez2004} scale, whilst bringing the observed and theoretical Li abundances closer, still failed to solve the lithium problem.

\begin{figure}[tbp]
\centering
\vbox{
			\psfig{file=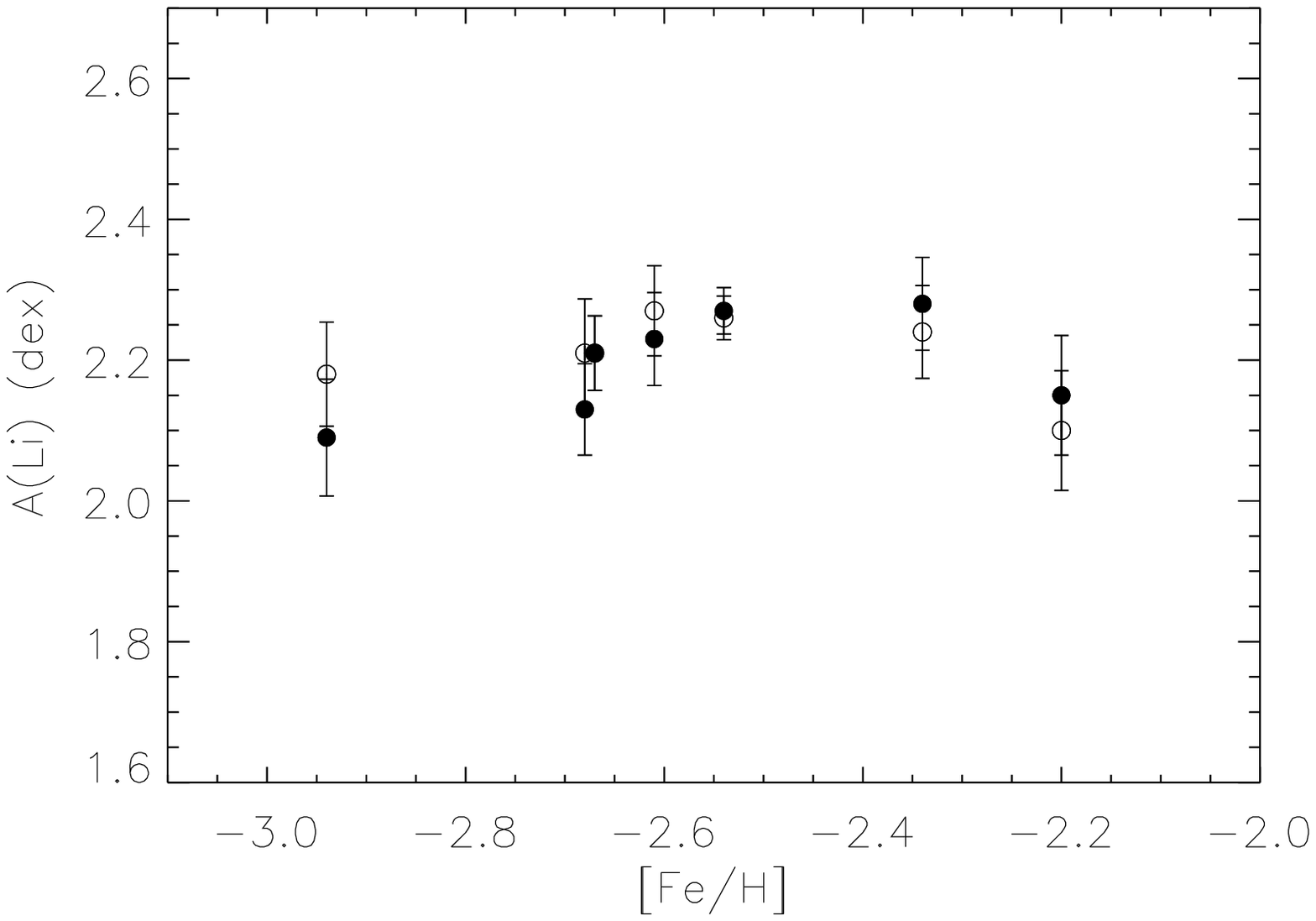,width=9cm,height=5.5cm}
			\psfig{file=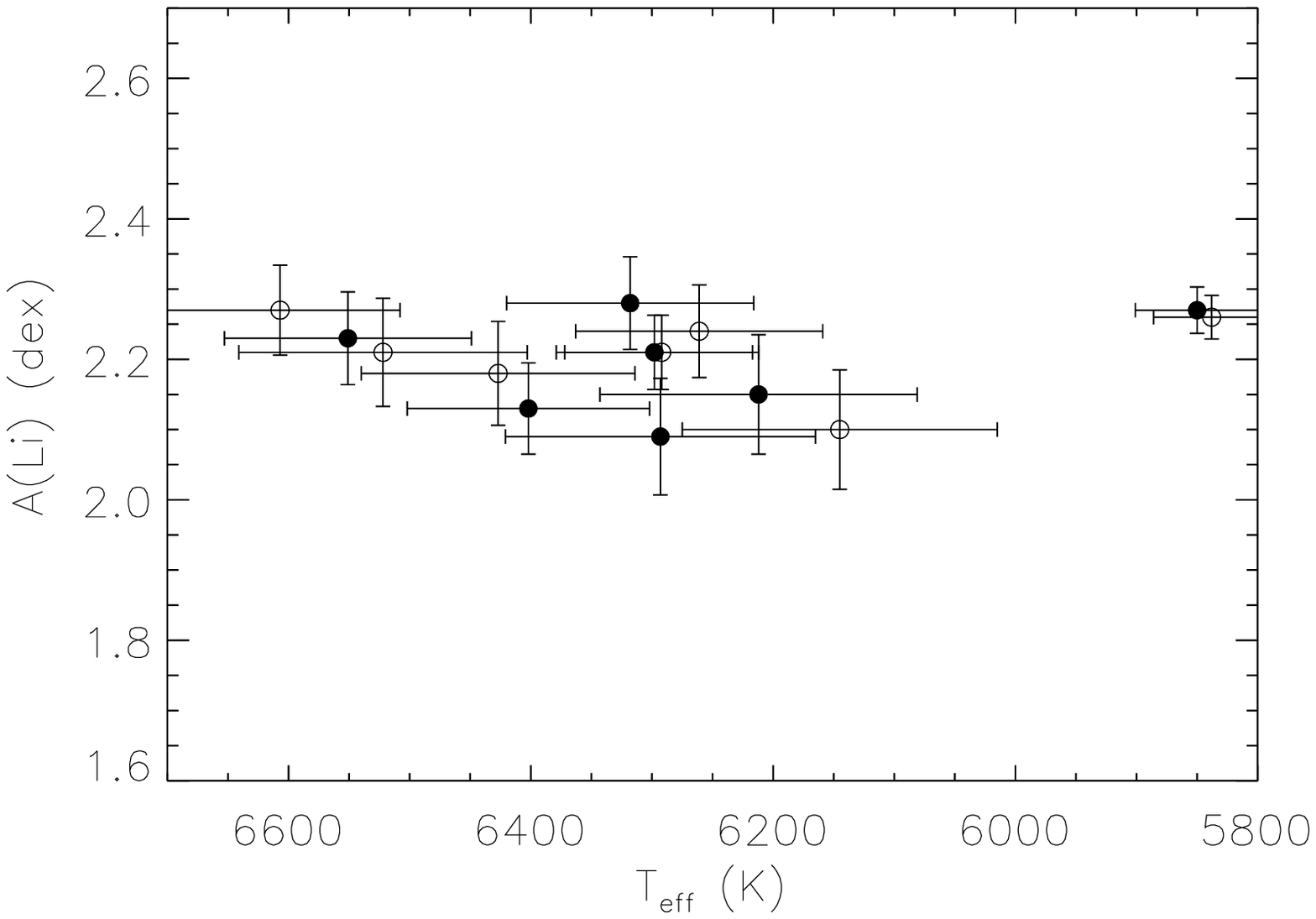,width=9cm,height=5.5cm}}
			\caption{Lithium abundance versus [Fe/H] (top panel) and $T_{\rm eff}$ (bottom panel) for $\rm S_{H}$ = 0 (filled circles) and 1 (open circles).}
			\label{Fig:met/teff-li}
\end{figure}

Fig. \ref{Fig:met/teff-li} shows the lithium abundances versus [Fe/H] and $T_{\rm eff}$, least squares fits have been performed for both sets of data. In the fit to metallicity we get the values:

\begin{equation}
A(\rm Li)=2.55(\pm0.31) + 0.137(\pm0.119)\rm [Fe/H]\\
\label{Lifit1}
\end{equation}
for $\rm S_{H}$ = 0 and
\begin{equation}
A(\rm Li)=2.04(\pm0.26) - 0.067(\pm0.102)\rm [Fe/H]\\
\label{Lifit2}
\end{equation}
for $\rm S_{H}$ = 1. For $\rm S_{H}$ = 0, we see a similar values to the coefficient of metallicity as \citet{Ryanetal1999}, whilst for $\rm S_{H}$ = 1, we have a value that is about half the size, and has a negative slope. However, our errors are much larger, due to the large random errors and small sample of stars, and therefore no statistically relevant trend can be deduced. For $T_{\rm eff}$, we get the equations:

\begin{equation}
A(\rm Li)=2.77(\pm0.91) - 0.00009(\pm0.00015)T_{\rm eff} \\
\end{equation}
for $\rm S_{H}$ = 0 and
\begin{equation}
A(\rm Li)=2.09(\pm0.63) + 0.00002(\pm0.00010)T_{\rm eff} \\
\end{equation}
for $\rm S_{H}$ = 1. Here we see no statistically relevant trend with $T_{\rm eff}$ for either $\rm S_{H}$ value.

We also perform the fit as described by \citet{Ryanetal2000}, such that:

\begin{equation}
Li/H = a^{'}+b^{'}Fe/Fe_{\sun}
\end{equation}
where $a^{'}$ measures the primordial abundance of Li and $b^{'}$ is a probe of galactic chemical evolution. For this fit, we obtain the primordial values of \element[][7]{Li}/\element[][]{H} = (1.47 $\pm$ 0.27)$\times10^{-10}$ for $\rm S_{H}$ = 0 and \element[][7]{Li}/\element[][]{H} = (1.80 $\pm$ 0.16)$\times10^{-10}$ for $\rm S_{H}$ = 1. Both of these values are far from the high value of \element[][7]{Li}/\element[][]{H} = $5.24^{+0.71}_{-0.62}\times10^{-10}$ \citep{Cyburt2008}from WMAP and BBN.

We see then that the addition of NLTE corrections has led to an increase in $T_{\rm eff}$ for most stars. This equates to an increase in $A$(Li) but it is still not high enough to reconcile the lithium problem. Through the efforts of \citet{Hosfordetal2009} and this study it is safe to conclude that systematic errors in the metal-poor $T_{\rm eff}$ scale are almost certainly not large enough to be the source of the $A$(Li) discrepancy between observation and WMAP + BBN predictions. This outcome lends strength to other possible explanations, such as processing in the stars, e.g. diffusion, processing in earlier generations of stars, and/or different BBN networks, or more exotic solutions requiring physics beyond the standard model.

It should be noted that while we have computed Fe lines in NLTE to constrain the temperature, our Li abundances are calculated from a grid of abundance versus equivalent width that was constructed under the assumptions of LTE, see Paper I for details. Several studies of the effects of NLTE Li line formation have been conducted. Two of these studies are those of \citet{Carlssonetal1994} and \citet{Lindetal2009}; they find Li abundance corrections of $\sim$ +0.013 -- +0.020 dex and $\sim$ +0.01 -- +0.03 dex respectively for the temperature, log g and [Fe/H] range in this study. Due to the very small size of these corrections we find the use of the LTE grid, combined with our NLTE effective temperatures, to be acceptable in determining Li abundances, and that the introduction of NLTE Li abundances will not significantly aid in solving the lithium problem.

\subsection{Implications of NLTE calculations for ionization balance and $\rm S_{H}$}
\label{sec:ImplicationsofNLTEcalculationsforionizationbalance}

Having discussed the effects of NLTE corrections on the $T_{\rm eff}$ scale and the lithium abundances, it is also of interest to note the effect on an aspect of abundance analysis, specifically ionization equilibrium often used in the determination of log $g$. We can also make a preliminary investigation into constraints we can place on the value of $\rm S_{H}$ from our results. 

It has been noted previously \citep{Gehrenetal2001} that \ion{Fe}{ii} lines are relatively unaffected by NLTE. In this work we have also found this to be the case with values for \ion{Fe}{ii} abundance corrections of order 0.01 dex. Our NLTE calculations induce a mean difference between $\Delta A$(\ion{Fe}{i}) and $\Delta A$(\ion{Fe}{ii}) of 0.39 dex for $\rm S_{H}$ = 0 and 0.27 dex for $\rm S_{H}$ = 1. Knowing that a 0.1 dex change in log g induces a difference of 0.05 dex between \ion{Fe}{i} and \ion{Fe}{ii} abundance, for there to be ionization balance, one would need a correction of $\sim$ + 0.8 dex and + 0.5 dex in log $g$ for $\rm S_{H}$ = 0 and 1 respectively. That is, due to overionization, forcing ionization balance for metal-poor dwarfs under LTE calculations would give log~$g$ values too low by 0.8 dex ($\rm S_{H}$ = 0) or 0.5 dex ($\rm S_{H}$ = 1). LTE calculations for HD140283 have occasionally yielded gravities lower than the {\sc hipparcos} gravity by $\sim$ 0.3 \citep[e.g.][]{Ryanetal1996b}, and for a selection of 13 halo main sequance turnoff stars with {\sc hipparcos} parallaxes Ryan et al. (2009 - in preperation) determine a mean difference of 0.2 in log $g$ compared to LTE ionization balance. These differences are less than what we compute for $\rm S_{H}$ = 1, and suggest that for the model atom we are using, the choice of $\rm S_{H}$ = 1 may underestimate the role of collisions with neutral hydrogen in diminishing the departures from LTE for Fe, i.e. that $\rm S_{H} > 1$. Whilst we have not attempted a detailed derivation of $\rm S_{H}$ by this method, \citet{Kornetal2003} has, arriving at a value of $\rm S_{H}$ = 3 based on the analysis of four halo stars and two others. Our results are broadly consistent with their conclusion.

\section{Conclusions}

We have discussed the processes of NLTE line formation of Fe lines. Here, we have shown the challenges posed by such calculations and the uncertainties that still arise, in particular due to the unknown magnitude of \ion{H}{} collisions. As there is at present no better theoretical or experimental description of the role of H collisions, one obvious next step would be to tie down the value of $\rm S_{H}$ for metal-poor stars, for example by forcing the equality of {\sc hipparcos} gravities and those determined by ionization equilibrium by changing $\rm S_{H}$ \citep{Kornetal2003}. For this reason we have discussed the effect of NLTE corrections on the ionization equilibrium and the magnitude of the effect on log $g$.

Six of the original program stars from Paper I have been analysed to calculate the effects of NLTE on the $T_{\rm eff}$ scale derived from \ion{Fe}{i} lines via excitation equilibrium. We have found that the effect of the correction is to cause an increase in $T_{\rm eff}$ ranging from 2 K to 150 K for $\rm S_{H}$ = 0 and 41 K to 122 K for $\rm S_{H}$ = 1. There is one exception; the star CD$-$33$^{\circ}$1173 has a negative correction ($-93$ K) for the $\rm S_{H}$ = 0 case. This may be due to the limited number of Fe lines available for this star, but also emphasises the intricacies of NLTE work which make it difficult to make reliable generalisations.

Our new temperatures have been compared to the photometric temperatures of \citet{Ryanetal1999}, the IRFM of \citet{MelendezRamirez2004}, and the Balmer line wing method of \citet{Asplundetal2006}. We find that the NLTE temperatures are hotter than \citet{Ryanetal1999} by an average of 132 K for $\rm S_{H}$ = 0 and 162 K for $\rm S_{H}$ = 1. Similar results are found when comparing against \citet{Asplundetal2006} with average differences of 76 K and 110 K for $\rm S_{H}$ = 0 and 1 respectively. The difference between our temperatures and the \citet{Asplundetal2006} temperatures may be removed if the Balmer line wing method suffers from NLTE effects \citep{Barklem2007}, or the effects of granulation are properly described. We find that even with NLTE corrections we are unable to match the high $T_{\rm eff}$'s of \citet{MelendezRamirez2004}. However, it has been acknowledged that their temperatures suffer from systematic errors (Mel$\rm \acute{e}$ndez 2009 - private communication) and a revision of their temperature scale is under way.

With our new $T_{\rm eff}$ scale, new Li abundances have been calculated. This has led to an increase of the mean Li abundance from \citet{Hosfordetal2009} to values of 2.19 dex with a scatter of 0.07 dex and 2.21 dex with a scatter of 0.06 dex for $\rm S_{H}$ = 0 and 1 respectively, both of which lie well below the value of 2.72 dex inferred from WMAP+BBN \citep{Cyburt2008}. This has shown that systematic errors in the $T_{\rm eff}$ scale of metal-poor stars are not the cause for the discrepency.

\begin{acknowledgements}
 
 The authors thank A. J. Korn for his discussions on the processes of NLTE and suggestions on how to proceed with this study. SGR \& AEGP gratefully acknowledge the support from the Royal Society under International Joint Project 2006/23 involving colleagues at Uppsala University. A.H. \& AEGP thank the STFC for its financial support to do this work. The work of KO was supported in part by DOE grant DE--FG02--94ER--40823 at the University of Minnesota.
      
\end{acknowledgements}

\nocite{*}
\bibliography{ref}
\bibliographystyle{aa}

\end{document}